%% file: TOP-21-013_temp.tex
\pdfoutput=1
\documentclass[11pt,twoside,a4paper,cmspaper,final,collab]{cms-tdr}
\def\svnVersion{817c11e}\def\svnDate{2024/04/06}\def\cmsCernNoTag{CERN-EP-2023-214}\def\cmsCernDate{\today}\def\cmsMessage{Published in Physical Review D as \href{http://dx.doi.org/10.1103/PhysRevD.109.072004}{\doi{10.1103/PhysRevD.109.072004}.}}
\begin{document}\cmsNoteHeader{TOP-21-013}

\newlength\cmsTabSkip\setlength{\cmsTabSkip}{1ex}
\newcommand{\tqg}{\ensuremath{\mathcal{\PQt\PQq\PGg}}\xspace}
\newcommand{\tug}{\ensuremath{\mathcal{\PQt\PQu\PGg}}\xspace}
\newcommand{\tcg}{\ensuremath{\mathcal{\PQt\PQc\PGg}}\xspace}
\newcommand{\BRtqg}{\ensuremath{\mathcal{B}(\PQt\to\PQq\PGg)}\xspace}
\newcommand{\BRtug}{\ensuremath{\mathcal{B}(\PQt\to\PQu\PGg)}\xspace}
\newcommand{\BRtcg}{\ensuremath{\mathcal{B}(\PQt\to\PQc\PGg)}\xspace}
\newcommand{\kappatqg}{\ensuremath{\kappa_{\tqg}}\xspace}
\newcommand{\kappatug}{\ensuremath{\kappa_{\tug}}\xspace}
\newcommand{\kappatcg}{\ensuremath{\kappa_{\tcg}}\xspace}
\newcommand{\mtop}{\ensuremath{m_{\PQt}}\xspace}
\newcommand{\mZ}{\ensuremath{m_{\PZ}}\xspace}
\newcommand{\pp}{\ensuremath{\Pp\Pp}\xspace}
\newcommand{\ttgamma}{\ensuremath{\ttbar\PGg}\xspace}
\newcommand{\tgamma}{\ensuremath{\PQt\PGg}\xspace}
\newcommand{\WWgamma}{\ensuremath{\PW\PW\PGg}\xspace}
\newcommand{\VVgamma}{\ensuremath{\PV\PV\PGg}\xspace}
\newcommand{\Zgamma}{\ensuremath{\PZ\PGg\text{+jets}}\xspace}
\newcommand{\Wgamma}{\ensuremath{\PW\PGg\text{+jets}}\xspace}
\newcommand{\tW}{\ensuremath{\PQt\PW}\xspace}
\newcommand{\Wjets}{\ensuremath{\PW\text{+jets}}\xspace}
\newcommand{\gammajets}{\ensuremath{\PGg\text{+jets}}\xspace}
\newcommand{\Zeejets}{\ensuremath{\PZ(\to\EE)\text{+jets}}\xspace}
\newcommand{\sigeta}{\ensuremath{\eta}\xspace}
\newcommand{\abseta}{\ensuremath{\abs{\eta}}\xspace}
\newcommand{\megamma}{\ensuremath{m_{\Pe\PGg}}\xspace}
\newcommand{\mlgamma}{\ensuremath{m_{\Pell\PGg}}\xspace}
\newcommand{\PQj}{{\HepParticle{j}{}{}}\xspace}
\newcommand{\Njet}{\ensuremath{N_{\PQj}}\xspace}
\newcommand{\Nbjet}{\ensuremath{N_{\PQb}}\xspace}
\newcommand{\DRlb}{\ensuremath{\DR(\Pell,\text{\PQb jet})}\xspace}
\newcommand{\DRtg}{\ensuremath{\DR(\PQt,\PGg)}\xspace}
\newcommand{\mjetgamma}{\ensuremath{m_{\text{jet+}\PGg}}\xspace}
\newcommand{\muR}{\ensuremath{\mu_{\mathrm{r}}}\xspace}
\newcommand{\muF}{\ensuremath{\mu_{\mathrm{f}}}\xspace}
\newcommand{\Leff}{\ensuremath{\mathcal{L}_{\text{eff}}}\xspace}
\newcommand{\Amu}{\ensuremath{A_\mu}\xspace}
\newcommand{\qnu}{\ensuremath{q_\nu}\xspace}
\newcommand{\sigmunu}{\ensuremath{\sigma^{\mu\nu}}\xspace}
\newcommand{\Cij}{\ensuremath{C_{ij}}\xspace}
\newcommand{\projL}{\ensuremath{P_{\mathrm{L}}}\xspace}
\newcommand{\projR}{\ensuremath{P_{\mathrm{R}}}\xspace}
\newcommand{\lLtqg}{\ensuremath{\lambda^{\mathrm{L}}_{\tqg}}\xspace}
\newcommand{\lRtqg}{\ensuremath{\lambda^{\mathrm{R}}_{\tqg}}\xspace}
\newcommand{\Deta}{\ensuremath{\Delta\eta}\xspace}
\newcommand{\Dphi}{\ensuremath{\Delta\phi}\xspace}
\newcommand{\ptpileup}{\ensuremath{\pt^{\text{pileup}}}\xspace}
\newcommand{\qfield}{\ensuremath{\overline{q}}\xspace}

\cmsNoteHeader{TOP-21-013}
\title{Search for flavor changing neutral current interactions of the top quark in final states with a photon and additional jets in proton-proton collisions at \texorpdfstring{$\sqrt{s}=13\TeV$}{sqrt(s) = 13 TeV}}

\date{\today}

\abstract{A search for the production of a top quark in association with a photon and additional jets via flavor changing neutral current interactions is presented. The analysis uses proton-proton collision data recorded by the CMS detector at a center-of-mass energy of 13\TeV, corresponding to an integrated luminosity of 138\fbinv. The search is performed by looking for processes where a single top quark is produced in association with a photon, or a pair of top quarks where one of the top quarks decays into a photon and an up or charm quark. Events with an electron or a muon, a photon, one or more jets, and missing transverse momentum are selected. Multivariate analysis techniques are used to discriminate signal and standard model background processes. No significant deviation is observed over the predicted background. Observed (expected) upper limits are set on the branching fractions of top quark decays: $\BRtug < 0.95\times10^{-5}$  ($1.20\times10^{-5}$) and $\BRtcg < 1.51\times10^{-5}$ ($1.54\times10^{-5}$) at 95\% confidence level, assuming a single nonzero coupling at a time. The obtained limit for \BRtug is similar to the current best limit, while the limit for \BRtcg is significantly tighter than previous results.}

\hypersetup{%
pdfauthor={CMS Collaboration},%
pdftitle={Search for flavor-changing neutral current interactions of the top quark in final states with a photon and additional jets in proton-proton collisions at sqrt(s)=13 TeV},%
pdfsubject={CMS},%
pdfkeywords={CMS, top quark, flavor-changing neutral current}}

\maketitle

\section{Introduction}

In the standard model (SM) of particle physics, flavor changing
neutral current (FCNC) interactions are not present at leading order (LO) and proceed through loop diagrams, which are strongly suppressed by the Glashow--Iliopoulos--Maiani (GIM)
mechanism~\cite{gim}. Because of this, the predicted branching fraction for a top quark decaying to an up or charm
quark and a photon \BRtqg, with $\PQq=\PQu$ or \PQc, is of the order of
$10^{-14}$~\cite{Aguilar-Saavedra:2004mfd, Aguilar-Saavedra:2002lwv, Larios:2006pb}.  While these branching fractions
are too small to be measured by current experiments, some extensions to the SM allow for significant enhancements to
them. For example, supersymmetry with $R$ parity violation, two-Higgs doublet models, and technicolor can enhance
\BRtqg by many orders of magnitude
compared to the SM value~\cite{Couture:1997ep,
Diaz:2001vj, Lu:2003yr}.  As a result, an observation of this signature would be a clear sign of physics beyond the
SM.

The \tqg FCNC interactions can be described by the effective field theory (EFT) framework in terms of
dimension-six operators added to the SM Lagrangian~\cite{AguilarSaavedra:2008zc}.
The most general effective Lagrangian up to dimension-six operators, \Leff, used to describe the
FCNC \tqg vertex has the following form:
\begin{linenomath}\begin{equation}\label{eq:lag}
\Leff = e \sum_{q=u,c}\kappatqg\qfield(\lLtqg\projL+\lRtqg\projR)\frac{i\sigmunu\qnu}{\mtop}t\Amu+\text{h.c.},
\end{equation}\end{linenomath}
where $e$ is the electric charge of the electron, \qnu is the
four-momentum of the photon, \mtop is the
top quark mass,
$\sigmunu=\frac{1}{2} [\gamma^\mu,\gamma^\nu]$,  \projL and \projR are
the left- and right-handed projection operators, $t$ is the top quark
field, \lLtqg and \lRtqg
are the fractions of the FCNC couplings for left- and right-handed chiralities, $\qfield = \overline{u}$ or $\overline{c}$ is the antiquark field, \Amu is
the electromagnetic field, and h.c.{} refers to the hermitian conjugate.
The strengths of the FCNC couplings are denoted by
\kappatqg, which are proportional to $\Cij/\Lambda^2$, where \Cij
are the Wilson coefficients defined in Ref.~\cite{AguilarSaavedra:2008zc}, $\Lambda$ is the energy scale of new
physics, and \kappatqg and \Cij are dimensionless coefficients.
In this analysis, no chirality is assumed for the FCNC interaction of
\tqg, and $\abs{\lLtqg}^2 + \abs{\lRtqg}^2 = 1$.
The values of \kappatqg vanish
in the SM at tree level.

\begin{figure}[!ht]
\centering
\includegraphics[width=0.45\textwidth]{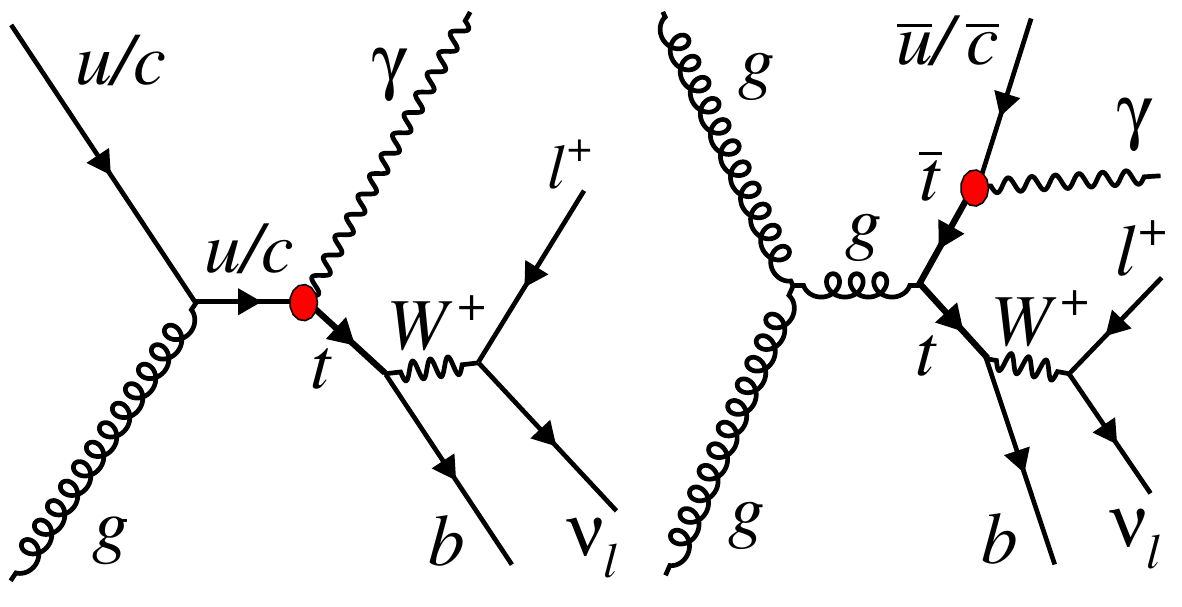}
\caption{LO Feynman diagrams for the production of a single top quark in association with
a photon (left), and the decay of a top antiquark to a photon and a light antiquark
in top quark pair production (right) via a \tqg FCNC, where $\PQq=\PQu$, \PQc. The leptonic decay of the \PW boson from the top quark decay
is included. The charge conjugate diagrams are also included.
The FCNC interaction vertex is marked as a filled red circle.}
\label{fig:feynman}
\end{figure}

Searches for the FCNC interactions \tug and
\tcg have been performed by several experiments, with no evidence of signal as yet~\cite{CDF:1997ydi, L3:2002hbp, H1:2009yuy, ZEUS:2011mya, Khachatryan:2015att, ATLAS:2022per}. The latest results of a search for \tqg FCNC couplings through single top
quark production in association with a photon by the CMS experiment
are $\BRtug < 1.3\times10^{-4}$, and
$\BRtcg < 1.7 \times 10^{-3}$~\cite{Khachatryan:2015att} at 95\% confidence level (\CL), based on proton-proton (\pp) collisions collected at a center-of-mass
energy of 8\TeV corresponding to an integrated luminosity of 19.8\fbinv.
The most stringent 95\% \CL limits to date have been obtained by the ATLAS
Collaboration at a center-of-mass energy of
13\TeV and corresponding to an integrated luminosity of 139\fbinv~\cite{ATLAS:2022per}. They report 95\% \CL upper limits of $\BRtug < 0.85\times10^{-5}$ ($1.2 \times 10^{-5}$) and $\BRtcg < 4.2\times10^{-5}$ ($4.5 \times 10^{-5}$) for left-handed (right-handed) chiralities, respectively.
This paper describes the search for \tug and \tcg FCNC couplings based on
the effective Lagrangian approach introduced in Eq.~\eqref{eq:lag}. The analysis considers
both the production of a single top quark with a photon, referred to as ST,
and the decay of a top quark to a photon and light-flavor quark (\PQu or \PQc)
in top quark pair production, referred to as TT.
The representative Feynman diagrams at LO for ST and TT are
shown in Fig.~\ref{fig:feynman}.

We focus on final states where the top quark decays into a \PW boson and a \PQb quark,
followed by the decay of the \PW boson to a neutrino and an electron or muon,
with electrons or muons from leptonic decays of tau leptons also being included.
The data for this analysis were collected by the CMS detector between 2016--2018 in \pp collisions at a center-of-mass
energy of 13\TeV, and correspond to an integrated luminosity of
138\fbinv. Tabulated results are provided in the HEPData record for this analysis~\cite{hepdata}.

\section{The CMS detector}

The central feature of the CMS apparatus is a superconducting solenoid of 6\unit{m} internal diameter,
providing a magnetic field of 3.8\unit{T}. Within the solenoid volume are a silicon pixel and strip
tracker, a lead tungstate crystal electromagnetic calorimeter (ECAL), and a brass and scintillator
hadron calorimeter (HCAL), each composed of a barrel and two endcap sections. Forward calorimeters
extend the pseudorapidity (\sigeta) coverage provided by the barrel and endcap
detectors, which improves the measurement of the imbalance in transverse
momentum (\pt). Muons are detected in
gas-ionization chambers embedded in the steel flux-return yoke outside the solenoid. In the
barrel section of the ECAL, an energy resolution of about 1\% is achieved for unconverted or
late-converting photons that have energies in the range of tens of \GeVns.
The remaining barrel photons have a resolution of about 1.3\% up to a pseudorapidity of $\abseta=1$,
rising to about 2.5\% at $\abseta=1.4$~\cite{Khachatryan:2015iwa}. The electron momentum is estimated by combining the energy measurement in the ECAL with the momentum measurement in the tracker. The momentum resolution for electrons with $\pt \approx 45\GeV$ from $\PZ\to\Pe\Pe$ decays ranges from 1.6 to 5\%. It is generally better in the barrel region than in the endcaps, and also depends on the bremsstrahlung energy emitted by the electron as it traverses the material in front of the ECAL~\cite{Sirunyan:2020ycc}. Muons are measured in the pseudorapidity range $\abseta<2.4$, with detection planes made using three technologies: drift tubes, cathode strip chambers, and resistive plate chambers. Matching muons to tracks measured in the silicon tracker results in a relative transverse momentum resolution, for muons with \pt up to 100\GeV, of 1\% in the barrel and 3\% in the endcaps~\cite{Sirunyan:2018fpa}.
Events of interest are selected using a two-tier trigger system. The
first level (L1), composed of
custom hardware processors, uses information from the calorimeters and muon detectors to
select events in a fixed time interval of less than 4\mus~\cite{CMS:2020cmk}. The second level, called the high-level
trigger, consists of a farm of processors running a version of the full event reconstruction software optimized for fast processing, and decreases the event rate from around 100\unit{kHz} to less than 1\unit{kHz} before
data storage~\cite{Khachatryan:2016bia}. A more detailed description of the CMS detector, together with a definition
of the coordinate system used and the relevant kinematic variables, can be found in Ref.~\cite{Chatrchyan:2008aa}.

\section{Signal and background simulation}

To account for the changes in conditions related to the CMS detector
over the three years in which the data were taken, the data from each
year are analyzed independently using the corresponding corrections and
calibrations.

Signal events and background processes are simulated for each of the three data-taking years using various
Monte Carlo (MC) generator packages. The signal samples corresponding to ST production associated with a photon,
and \ttbar production with an FCNC decay of one of the top quarks, are generated at LO
with \MGvATNLO 2.4.2~\cite{Alwall:2014hca}.
Based on the cross section calculation with the \textsc{top++} 2.0
program~\cite{Czakon:2011xx} at next-to-next-to-LO in quantum chromodynamics (QCD)
including soft-gluon resummation to next-to-next-to-leading-logarithmic order, the top quark pair production cross section is taken as 832\unit{pb}.
Two signal scenarios are generated: $(\kappatug=0.1, \kappatcg=0.0)$ and
$(\kappatug=0.0, \kappatcg =0.1)$, with corresponding LO ST cross sections of 0.707 and 0.100\unit{pb}, respectively, and TT cross section times FCNC branching fraction of 1.367\unit{pb} for both scenarios. The $\sigma_{\text{NLO}}/\sigma_{\text{LO}}$ $K$ factors of 1.37 and 1.41~\cite{Zhang:2011gh} are used to account for next-to-LO (NLO) QCD corrections for \tug and \tcg couplings, respectively. Since the final state kinematical distributions are independent of the FCNC \tqg couplings, we only need to generate ST and TT signal samples for the two scenarios described above (and scale them as necessary).

The \MGvATNLO 2.4.2 package is used to generate most of the background processes
including \ttgamma, \tgamma,  Drell--Yan, \WWgamma, \Zgamma, \Wgamma, and $\PV\PV$, where $\PV = \PW$, \PZ,
at NLO.
The single top quark production processes ($t$-, $s$-, \tW-channel) and top quark pair production
 are generated at NLO with \POWHEG 2.0~\cite{Nason:2004rx, Frixione:2007vw, Alioli:2010xd}.
The simulation of signal and all background events is interfaced with \PYTHIA 8.2~\cite{Sjostrand:2014zea} for parton showering,
fragmentation, and hadronization. The underlying event is modeled using \PYTHIA 8.2 with the
\textsc{CUETP8M1}~\cite{Khachatryan:2015pea} and \textsc{CP5}~\cite{Sirunyan:2019dfx} tunes for 2016 and 2017--2018 data, respectively. The parton distribution functions (PDFs) that are used for 2016 simulation are NNPDF 3.0~\cite{NNPDF:2014otw} and for 2017/2018 simulation are NNPDF 3.1~\cite{Ball:2017nwa}.

The CMS detector response for both signal and background processes is modeled with the \GEANTfour program~\cite{Agostinelli:2002hh}.
Simulated minimum-bias interactions in nearby or in the same \pp bunch crossing,
referred to as ``pileup'', are included.
Events in the simulation are reweighted to reproduce the pileup distribution observed in the data~\cite{Sirunyan:2018nqx}.

\section{Event reconstruction}
 The particle-flow (PF) algorithm is used to reconstruct and identify the particles (photons, electrons, muons, charged and neutral hadrons)
 produced in the event by combining the information from various subdetectors~\cite{Sirunyan:2017ulk}.
The primary vertex (PV) is taken to be the vertex corresponding to the hardest scattering in the event, evaluated using tracking information alone, as described in Section 9.4.1 of Ref.~\cite{Contardo:2015bmq}.
In this analysis, events of interest contain one isolated charged lepton (either a muon or an electron),
an isolated photon, at least one jet, and missing transverse momentum.
Quality criteria are applied to improve the purity
of events containing genuine leptons and photons and to help reject
candidates that originate from misidentified jets or from pileup events.

Photon candidates are identified according to the presence of an energy deposit in the ECAL
with no associated charged-particle tracks matched with this deposit~\cite{Sirunyan:2020ycc}.
The photon energy is obtained from the ECAL measurement. Corrections to account for
the zero suppression and energy scale are applied in both simulation and data~\cite{Sirunyan:2020ycc}.
The reconstructed photons must satisfy quality criteria
based on the following quantities:
the relative amount of deposited energy in the ECAL and HCAL;
a ``shower shape'' variable quantifying the lateral development of the shower~\cite{Sirunyan:2020ycc};
separate isolation variables for charged hadrons, neutral hadrons, and photons;
and a variable that quantifies whether the photon candidate is consistent with a hit in the pixel detector.

Electron reconstruction relies on a match between the clusters in
the ECAL and the tracks in the tracker. The energy of an
electron is estimated from the combination of electron momentum as
determined by the tracker at the PV and the energy of the
corresponding cluster in the ECAL including the energy sum of all the
bremsstrahlung photons compatible with the electron track~\cite{Sirunyan:2020ycc}.
Electrons are required to satisfy identification criteria related to the track
impact parameters with respect to the PV,
the matching between the ECAL cluster and the associated track,
the ratio of hadronic to electromagnetic energy deposits, and the shower shape in the ECAL.
A veto on electrons originating from photon conversions is
applied~\cite{Sirunyan:2020ycc}.

Muon reconstruction is based on the combination of information from the tracker and the muon system~\cite{Sirunyan:2018fpa}.
A global track fit is produced using the combined information,
and the muon momentum is obtained from the track curvature.
Identification criteria are based on the impact parameters with respect to the PV,
the number of hits in the muon systems and the tracker,
the number of matched muon detector planes,
as well as the fit quality of the combined muon track.

The photon and charged lepton candidates are required to be
isolated from other particles in an event.
For electrons and muons (\Pell), a relative isolation variable is defined as
\ifthenelse{\boolean{cms@external}}{
    \begin{linenomath}\begin{multline}
        I^{\Pell}_{\text{rel}} = \frac{1}{\pt^{\Pell}}  \Bigg[
          \sum_{\text{charged}} \pt  \\
          +\max \Bigg(0, \Big[\sum_{\text{neutral}} \pt + \sum_{\text{photons}} \pt\Big] - \ptpileup\Bigg)     \Bigg],
        \end{multline}\end{linenomath}
}{
\begin{linenomath}\begin{equation}
I^{\Pell}_{\text{rel}} = \frac{1}{\pt^{\Pell}}  \Bigg[
  \sum_{\text{charged}} \pt  +\max \Bigg(0, \Big[\sum_{\text{neutral}} \pt + \sum_{\text{photons}} \pt\Big] - \ptpileup\Bigg)     \Bigg],
\end{equation}\end{linenomath}
}
where the sums run over the \pt of charged and neutral hadrons, as well as the
photons, in a cone of $\DR < 0.3$ (0.4) centered on the electron (muon),
where $\DR = \sqrt{\smash[b]{(\Deta)^2+(\Dphi)^2}}$. Here,
\Deta is the difference in pseudorapidity and \Dphi is
the difference in azimuthal angle between the direction of the lepton and the other
particles inside the isolation cone.
To reduce the pileup effects, only charged hadrons compatible with originating at the PV are taken into account.
For the sums corresponding to photons and neutral hadrons, an estimate of the expected
pileup contribution (\ptpileup) is subtracted. For photons, three separate isolation
variables, corresponding to charged hadrons, neutral hadrons, and
other photons, are defined within a cone of $\DR < 0.3$~\cite{Sirunyan:2020ycc}. The
effect of pileup in each variable is subtracted~\cite{CMS:2015xaf}.

The muon and electron requirements, including the identification and
isolations criteria mentioned above, describe the ``tight'' selection, and are intended to efficiently
select leptons originating from the decay of gauge bosons~\cite{Sirunyan:2020ycc,Sirunyan:2018fpa}.  A less stringent set of requirements, referred to as the
``loose'' selection, is also used in the analysis for the purpose of vetoing events with additional leptons and to
define a control region (CR) enriched in non-genuine leptons.  The tight leptons are a subset of the loose leptons.

Jets are clustered from the PF candidates using the infrared- and collinear-safe anti-\kt
algorithm~\cite{Cacciari:2008gp, Cacciari:2011ma} with a distance parameter of $\DR=0.4$.
To correct for the contributions from pileup, charged particles
originating from vertices other than the PV are discarded and a
correction is applied to the neutral contributions.
Jet energy corrections are derived from simulation to, on average, equal the summed energy of the truth-level particles in the jet. In situ measurements of the momentum balance in dijet,
{\PGg}+jet, {\PZ}+jet, and multijet events are
used to account for any residual differences in the jet energy scale between data and simulation~\cite{Khachatryan:2016kdb}. The jet energy
resolution (JER) is typically 15--20\% at 30\GeV, 10\% at 100\GeV, and 5\% at 1\TeV~\cite{Khachatryan:2016kdb}.
Additional selection criteria are applied to each jet to remove jets potentially dominated by anomalous contributions from various subdetector components or reconstruction failures~\cite{CMS:2017wyc}.

The missing transverse momentum vector (\ptvecmiss) is computed
as the negative vector \pt sum of all the PF
candidates, and its magnitude
is denoted as \ptmiss~\cite{Sirunyan:2019kia}.
The \ptvecmiss is modified to account for
corrections to the reconstructed energy scale of jets.

\section{Event selection and analysis strategy}

Events are collected with single-lepton triggers, with minimum
\pt thresholds for electrons of 27\GeV in 2016 and 32\GeV in 2017
and 2018, and for muons of 24\GeV in 2016 and 27\GeV in 2017 and 2018.

Events are selected by requiring the presence of either a tight electron with $\pt>35\GeV$ and $\abseta<2.5$
or a tight muon with $\pt>30\GeV$ and $\abseta<2.4$ that matches the triggering lepton.
Events with only a tight electron candidate in the ECAL barrel-endcap
transition region of $1.44<\abseta<1.57$  are rejected because the reconstruction of electrons in
this region is suboptimal. Events are discarded if they contain an additional loose electron with $\pt>20\GeV$, or loose muon with $\pt>15\GeV$.
 Photons are required to have $\pt>30\GeV$ and  $\abseta<1.44$, and pass the medium identification and isolation requirements~\cite{Sirunyan:2020ycc}. Photons in the endcap region $\abseta>1.57$
 are not considered as the signal purity is lower in this region.

For both electron and muon channels, we require $\ptmiss>30\GeV$ to reduce the background from events without neutrinos.
In the electron channel, due to the higher probability of electrons to
be misidentified as a photon, a veto on events with $81<\megamma<101\GeV$ is applied to suppress the contribution of $\PZ\to\EE$ events.

Events are required to have at least one reconstructed jet with $\pt>30\GeV$ and $\abseta<2.7$ or
$\pt>60\GeV$ and $2.7<\abseta<3.0$.  Higher \pt is required in the forward region to mitigate an excess of jets caused by noise in the endcap of ECAL.
The selected jets are required to be separated from the selected lepton by $\DR > 0.4$.
The \textsc{DeepCSV} tagging algorithm~\cite{CMS:2017wtu} is used to identify jets originating from the hadronization
of \PQb quarks.  The chosen selection requirements correspond to an efficiency
of around 70\% in identifying \PQb jets and to a misidentification
rate of 12\% for \PQc quark jets, and 1\% for light-quark and gluon
jets. The region over which the b jets can be identified increased from $\abseta < 2.4$ in 2016 to $\abseta < 2.5$ in 2017--2018 following an upgrade to the
pixel detector~\cite{pixel_upgrade}.
The top quark is reconstructed using the lepton, the \PQb-tagged jet,
 and \ptvecmiss, where the neutrino \pt is assumed to be equal to \ptmiss.
 The longitudinal component of the neutrino momentum is obtained by constraining the invariant mass of the lepton and the neutrino to the nominal \PW boson mass~\cite{PDG2022},
 and the top quark momentum is reconstructed from the combination of the momenta of the reconstructed \PW boson
 and the \PQb jet~\cite{D0:2009isq, CDF:2009itk, Chatrchyan:2011vp}.

In addition to the above requirements, photons are required to be well separated from the
selected lepton and jets by $\DR > 0.5$. This requirement helps to avoid distortion of the photon energy measurement due to the presence of close-by jets. Furthermore, it reduces the contributions of photons emitted from top quark decay products.

Two statistically independent signal regions (SRs) are defined for each lepton channel:
\begin{itemize}
\item SR1: exactly one \PQb-tagged jet and no additional jet
    ($\Njet = 1$ and $\Nbjet = 1$). This region targets the signal from FCNC ST quark production with a photon.
\item SR2:\ at least two jets, exactly one of which is \PQb-tagged
    ($\Njet \geq 2$ and $\Nbjet= 1$).
This region targets TT events with the FCNC decay of one of the top quarks.
\end{itemize}

\section{Background estimation}
\label{sec:bkg}

Several processes with identical or similar final states to the signal are sources of background for this analysis.
We distinguish four different sources of background: nonprompt-photon, nonprompt-lepton, misidentified-photon, and
irreducible backgrounds.  Nonprompt-photon backgrounds come from events in which a jet is misidentified as a photon.  This is usually the
result of a high-\pt \PGpz or \PGh meson in the jet decaying to a pair of photons.  Nonprompt-lepton backgrounds are
from events in which the selected lepton is from the decay of a hadron.  Misidentified photon backgrounds originate from
events in which an electron is misidentified as a photon.  Irreducible backgrounds are events in which a genuine photon
and charged lepton originate from the \pp collision vertex. Expected
and observed event yields for all background processes are presented in Table~\ref{tab:eventYield},
where all background contributions are predicted from data except for \tgamma and \VVgamma. The estimated backgrounds are described in the rest of this section.

\subsection{Nonprompt-photon background}

Nonprompt-photon events arise primarily from \Wjets and \ttbar events when the jets have
a high fraction of their energy deposited in the ECAL, and thus mimic prompt photons.  Because the probability for jets
to satisfy the photon identification criteria is not well modeled by the simulation, this source of background is estimated
from data.

{\tolerance=800
The SR photon requirements include selections on the charged-hadron isolation, neutral-hadron isolation, and photon isolation, as well as requirements on the shower shape and the ratio of hadronic to electromagnetic energy.
These five requirements are placed at the ``medium operating point'' of \cite{Sirunyan:2020ycc}
with average signal efficiency of 80\%.
Each of these variables also has a ``loose operating point'' with
average signal efficiency of 90\%,
which is less stringent than the medium one.
\par}

We begin by using the matrix ``ABCD'' method, with the two axes being the charged-hadron isolation variable and the
shower shape variable, which are found to be uncorrelated. The A region is the SR, containing events in which the photon candidate passes all medium
selections.  The B, C, and D regions are similar to the SR except for
these two mentioned variables. The B and C regions contain events where either the charged-hadron isolation or the shower shape
variable fails the loose selection, and the other one passes the medium selection. The D
region is constructed from events that fail the loose selection of both the charged-hadron isolation and shower shape variables. An estimate of the number of nonprompt-photon events in region A (the SR)
can be obtained from $N_\text{B}N_\text{C}/N_\text{D}$.

In addition to the normalization of the nonprompt-photon background in the signal region, a full set of background distributions is needed.  Therefore, we also
use a CR enriched in nonprompt-photons called the photon-like jet (PLJ) sample.
The PLJ sample is created by selecting events in which the photon candidate fails exactly one of the five selection criteria at the loose
operating point while passing the other criteria at the medium operating point.

We then define an extrapolation factor:
\begin{linenomath}\begin{equation}
    \text{EF}(\pt) = \frac{N_{\text{B}(\pt)} N_{\text{C}(\pt)}}{N_{\text{D}(\pt)} N_{\text{PLJ}(\pt)}},
\end{equation}\end{linenomath}
which is calculated in \pt bins, as the fraction of nonprompt
photons is known to depend on \pt.  All of the events in the PLJ sample are weighted
by $\text{EF}(\pt)$ to obtain the estimated distribution of nonprompt-photon events in the SR.  The estimation is performed separately for each
lepton channel and for each data-taking year.

The method is validated using simulation samples. In the first step,
the extrapolation factors are recalculated using simulated
samples. Then the contribution of the \Wjets sample, which is one
of the main sources of nonprompt-photon background in the SR, is compared to the value estimated by the method described above for this sample. The comparison shows an agreement between 10--15\% depending on the years and lepton channels.

\subsection{Nonprompt-lepton background}
\label{sec:nonpromptlep}

The nonprompt-lepton background contribution is estimated from data using the ``tight-to-loose ratio''
method~\cite{CMS:2011mlu}.  A dijet CR enriched in nonprompt-lepton events is selected by requiring exactly
one loose lepton, exactly one jet with $\pt>30\GeV$ and $\abseta<2.5$, $\ptmiss<30\GeV$, $\DR(\Pell,\text{jet})>0.3$, and a transverse mass of the lepton and \ptmiss less than 20\GeV.  The last criterion is
used to suppress \Wjets events.  The \ptmiss requirement ensures that this CR is disjoint from the SRs.  The contribution of events with prompt leptons is estimated from simulation and subtracted from the sample.

The nonprompt rate (NPR) is calculated from this sample as the fraction of events in which the selected lepton passes
the tight selection.  This rate is measured in bins of lepton \pt and \sigeta and converted to a weight $w = \text{NPR}/(1-\text{NPR})$.
A second CR is defined by requiring all of the SR criteria, except that the lepton
requirement is changed to select leptons that pass the loose requirement and fail the tight requirement.  The
contribution of events with prompt leptons is estimated from simulation and subtracted from this sample.  The
weight $w(\pt^{\Pell},\eta^{\Pell})$ is applied to all events in this CR to estimate the number of nonprompt-lepton events in the SR.

The method is validated in simulation using a \gammajets sample, which is the dominant source of the nonprompt-lepton
background.  The contribution of \gammajets events in the SR is compared to the estimate from the above
method applied to the same simulated events and found to deviate by no more than 20\%.

\subsection{Misidentified-photon background}

The misidentified-photon background arises from events in which an electron is misidentified as a photon.  This can
originate in both channels from \ttbar, diboson, and single top
quark \tW-channel production.  In addition,
\Zeejets events contribute to the electron channel.

The normalization for this background source is estimated using two CRs.  The first CR has
the same criteria as for the signal region, except that only the electron channel is considered and the invariant mass
of the electron and photon is required to be in the range of $\abs{\megamma - \mZ} < 10\GeV$ (the inverse of the
veto used in the SR definition).  The second CR also uses the SR criteria, except instead
of requiring a photon and an electron, we require two electrons passing the
tight selection with an invariant mass of $\abs{m_{\Pe\Pe} - \mZ} < 10\GeV$.  Both regions select
\Zeejets events, with the second CR containing two correctly identified electrons and the
first CR containing one electron misidentified as a photon.  The ratio of the number of events in the first CR
to the number of events in the second CR is calculated for both data and MC simulation.  From these two ratios,
the double ratio is calculated as:
\begin{linenomath}\begin{equation}
    \text{SF}(\Pe\to\PGg)= \frac{N_{\Pe\PGg}(\text{data})/N_{\Pe\Pe}(\text{data})}{N_{\Pe\PGg}(\text{MC})/N_{\Pe\Pe}(\text{MC})},
\end{equation}\end{linenomath}
which is applied as a scale factor to simulated background events in each SR in which the reconstructed
photon is matched to a generated electron.

\subsection{Irreducible backgrounds}

Irreducible backgrounds arise from \ttgamma, \Wgamma, \Zgamma, \tgamma, and \VVgamma
processes.  The shapes of these background sources are obtained from simulation.  The normalizations for the \tgamma
and \VVgamma processes are obtained from their respective NLO cross sections, while the normalizations for the other processes are constrained from data using CRs.  The \ttgamma CR is constructed with the
same requirements as in the SRs, but with $\Njet \geq 2$ and $\Nbjet \geq 2$.  A single-bin fit
is made in this CR to obtain the correction factor that can be used in
the SRs to correct the estimated number of \ttgamma events from MC.  The \Wgamma and \Zgamma CRs use the same criteria as the SRs but with
$\Njet \geq 1$ and $\Nbjet = 0$ and are distinguished from each other by requiring the invariant mass of
the lepton and photon to satisfy, for the electron channel, $\mlgamma<\mZ-10\GeV$ for the \Zgamma CR and $\mlgamma>\mZ+10\GeV$ for the \Wgamma CR, and for the muon channel, $\mlgamma<\mZ$ for the \Zgamma CR and $\mlgamma>\mZ$ for the \Wgamma CR. The reason that \Zgamma
events are populated in the low \mlgamma region is that in many cases the selected photon is radiated off the lepton from the \PZ boson decay with high angular separation. The normalization of each process in the SR is corrected by scale factors obtained from a
simultaneous fit of the \mlgamma distribution in the two CRs.  For all three CRs, the
background from other sources is estimated using the methods described
above for nonprompt-photon, nonprompt-lepton, and misidentified-photon
backgrounds, and from simulation for the remaining sources.

\subsection{Summary of the backgrounds}

Event yields in data and MC for all background processes are presented
in Table~\ref{tab:eventYield}. The irreducible backgrounds from
\ttgamma, \Wgamma, and \Zgamma are normalized with the control regions. The remaining irreducible backgrounds (\tgamma and \VVgamma) are estimated from simulation. The nonprompt-photon, nonprompt-lepton, and misidentifed-photon backgrounds are estimated from data.

\begin{table*}[!htp]
\centering
\topcaption{Estimated background yields and observed event counts for the
electron and muon channels in the signal regions SR1 and SR2.  The uncertainties
are the statistical and systematic contributions summed in quadrature.}
\label{tab:eventYield}
\renewcommand{\arraystretch}{1.2}
\begin{scotch}{lr@{${}\pm{}$}lr@{${}\pm{}$}lr@{${}\pm{}$}lr@{${}\pm{}$}l}
Process                           & \multicolumn{2}{c}{SR1 (\Pe channel)} & \multicolumn{2}{c}{SR2 (\Pe channel)} & \multicolumn{2}{c}{SR1 (\PGm channel)} & \multicolumn{2}{c}{SR2 (\PGm channel)} \\ \hline
\ttgamma                   & 360  &  76 & 4270  & 900   & 420  & 72  & 5020  & 850  \\
\Wgamma                     & 850  & 130 & 1170  & 210   & 757  & 76  & 970   & 110  \\
\Zgamma                     & 366  & 66  & 550   & 150   & 900  & 120 & 900   & 210  \\
\tgamma                        & 288  & 85  & 1030  & 300   & 319  & 93  & 1140  & 340  \\
\VVgamma                    & 5    & 3   & 37    & 17    & 6    & 4   &  38   & 16   \\
$\text{jet}\to \PGg$      & 532  & 97  & 1970  & 360   & 579  & 97  & 2100  & 350  \\
$\text{jet} \to \Pe/\PGm$ & 650  & 130 & 1100  & 220   & 570  & 110 &  790  & 160  \\
$\Pe \to \PGg$            & 1050 & 280 & 3090  & 760   & 800  & 210 & 2930  & 730  \\ [\cmsTabSkip]
Total                             & 4100 & 370 & 13200 & 1300  & 4350 & 320 & 13900 & 1300 \\[\cmsTabSkip]
Data                              & \multicolumn{2}{l}{4308} & \multicolumn{2}{l}{13166} & \multicolumn{2}{l}{4252} & \multicolumn{2}{l}{13530} \\
\end{scotch}
\end{table*}

\section{Discrimination of signal and background}

{\tolerance=800
To distinguish the potential FCNC signal from backgrounds, a number of discriminant variables
are combined into a multivariate classification based on boosted decision trees (BDTs)~\cite{tmva}.
Separate BDT classifiers are trained for the two signal scenarios (\tug and \tcg),
the two SRs (SR1 and SR2), and the two channels (electron and muon), for a total of eight BDTs.
The BDT input variables include photon \pt and \sigeta,  lepton
\sigeta, reconstructed top quark mass ($m_{\Pell\PGn\PQb}$),
transverse mass of the \PW boson defined as $m_{\PW\mathrm{T}} = \sqrt{\smash[b]{2\pt^{\Pell}\ptmiss (1-\cos\Dphi)}}$ where \Dphi is the difference in azimuthal angle between the lepton and \ptvecmiss, invariant mass of the
photon and non-\PQb jet (\mjetgamma),
\ptmiss, $\DR(\Pell, \PGg)$, \DRtg,
jet multiplicity, \DRlb, $\DR(\text{\PQb jet}, \PGg)$.
In addition, for the \tug signal, the lepton charge
is added as an input variable, to provide sensitivity to the asymmetry
between top quark and top antiquark production~\cite{Khatibi:2014via}, which arises from different PDFs for the up quark and up antiquark in the colliding protons.
Simulated \tug and \tcg signal processes are used in conjunction with simulated \Wgamma, \ttgamma, \Zgamma, and
\tgamma background events to train the BDTs. The
four most discriminating BDT variables for SR1 are photon \pt,
\DRlb, reconstructed top quark mass, and
\DRtg, and for the SR2 they are
\mjetgamma, photon \pt, reconstructed top quark mass, and
\DRlb.
\par}

\begin{figure*}[!p]
\centering
\includegraphics[width=0.40\textwidth]{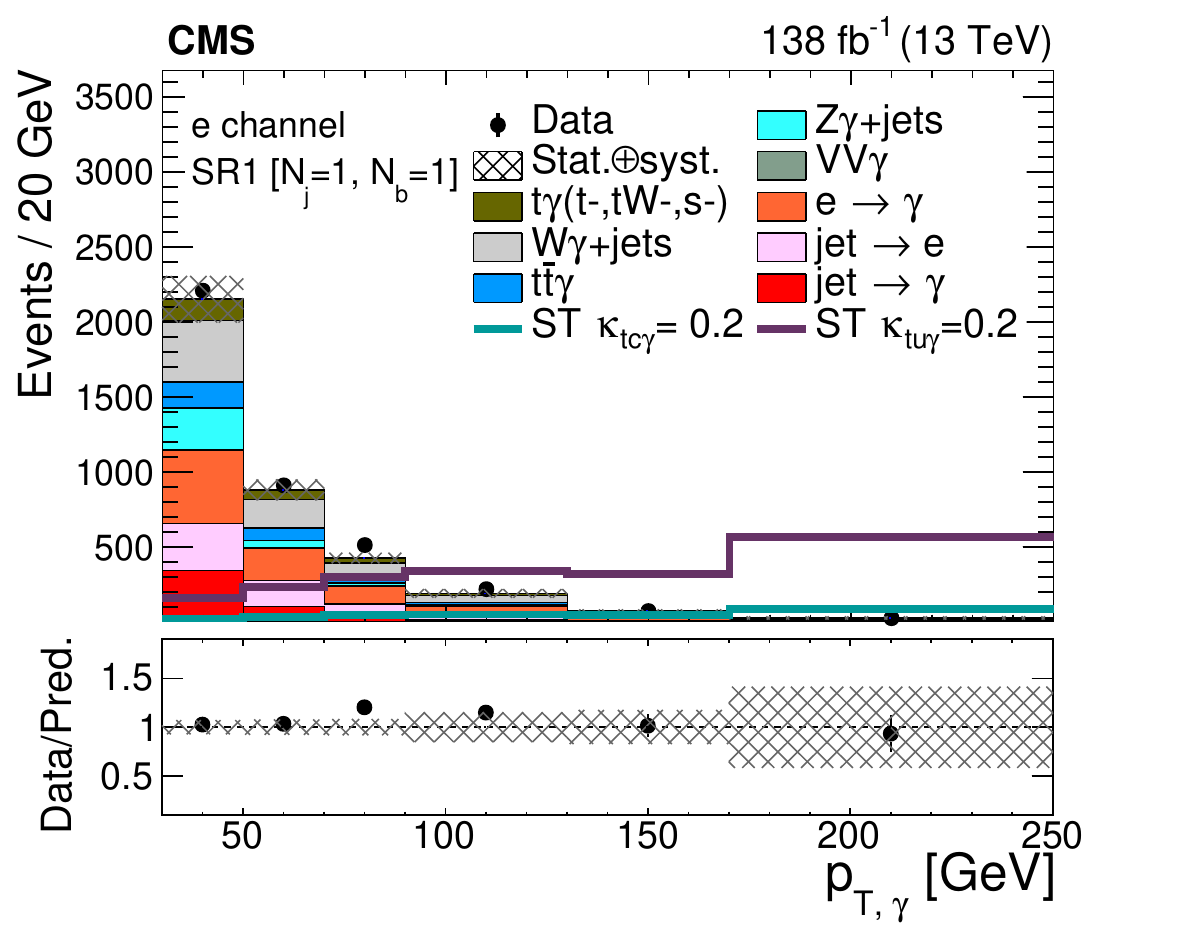}
\includegraphics[width=0.40\textwidth]{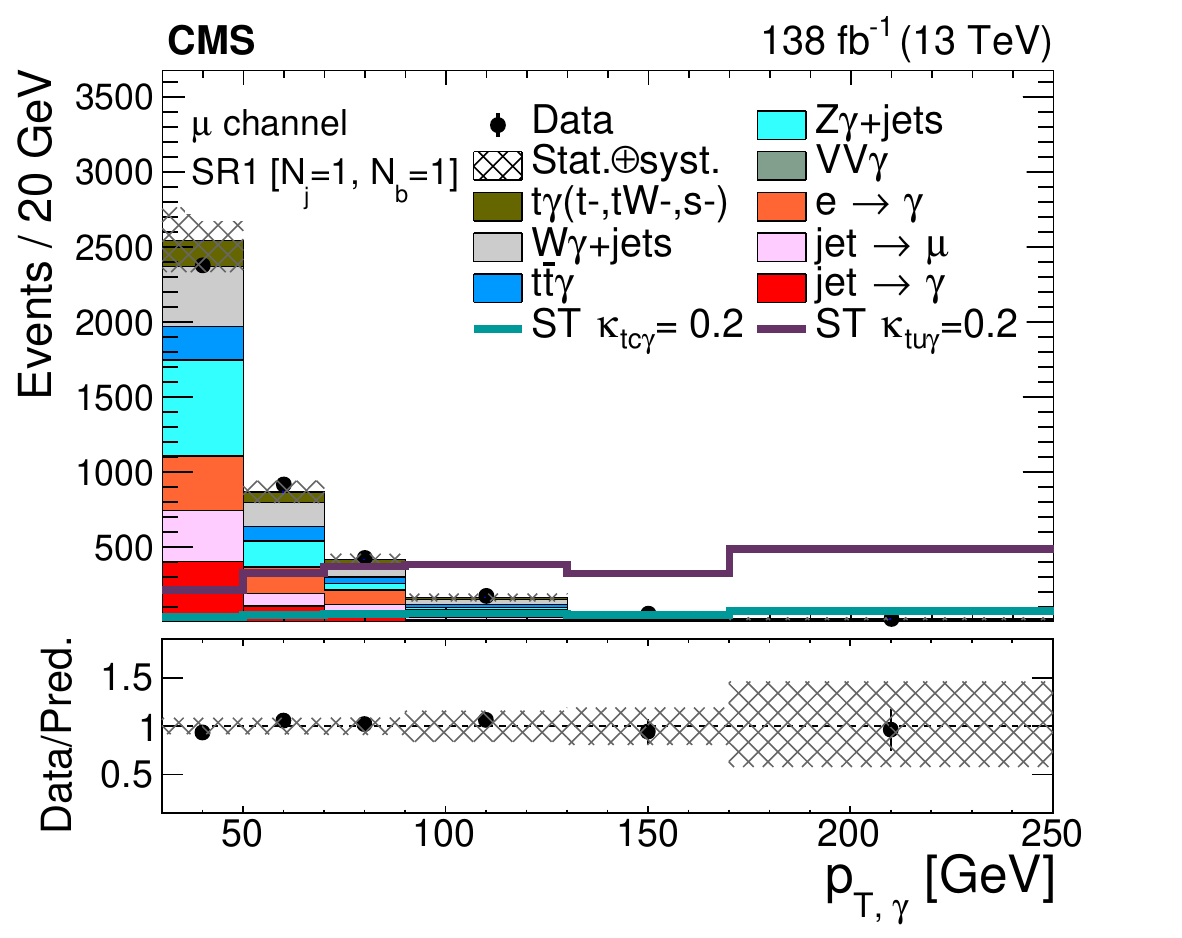}
\includegraphics[width=0.40\textwidth]{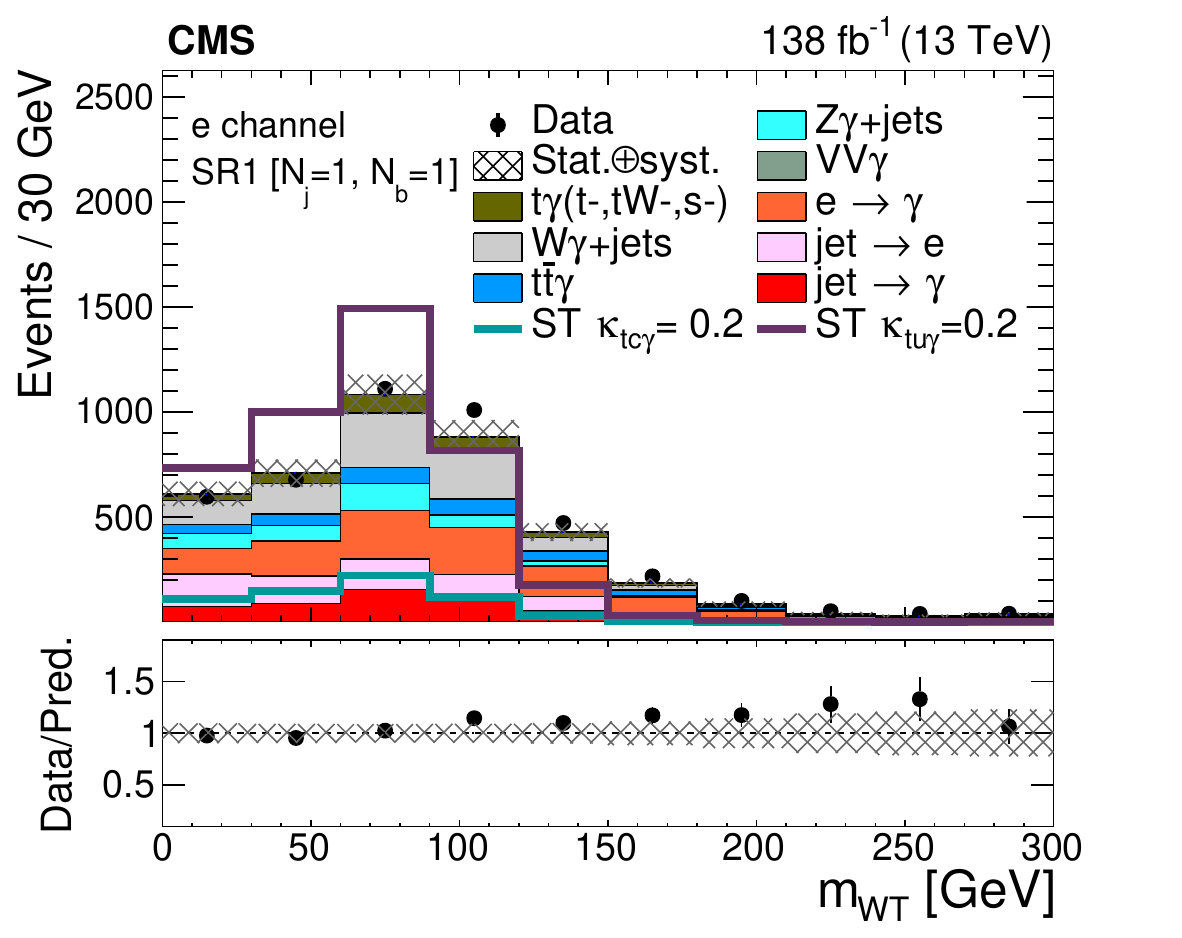}
\includegraphics[width=0.40\textwidth]{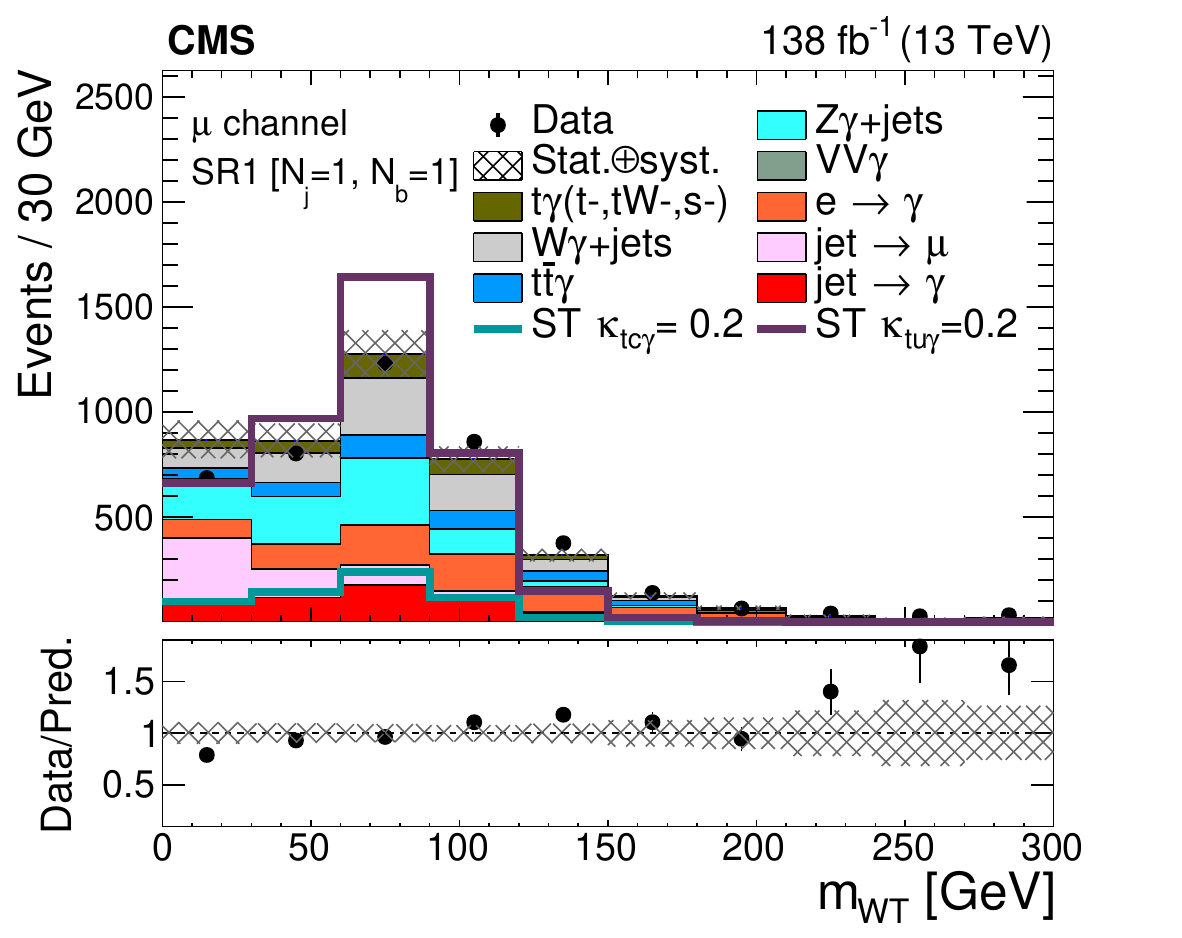}
\includegraphics[width=0.40\textwidth]{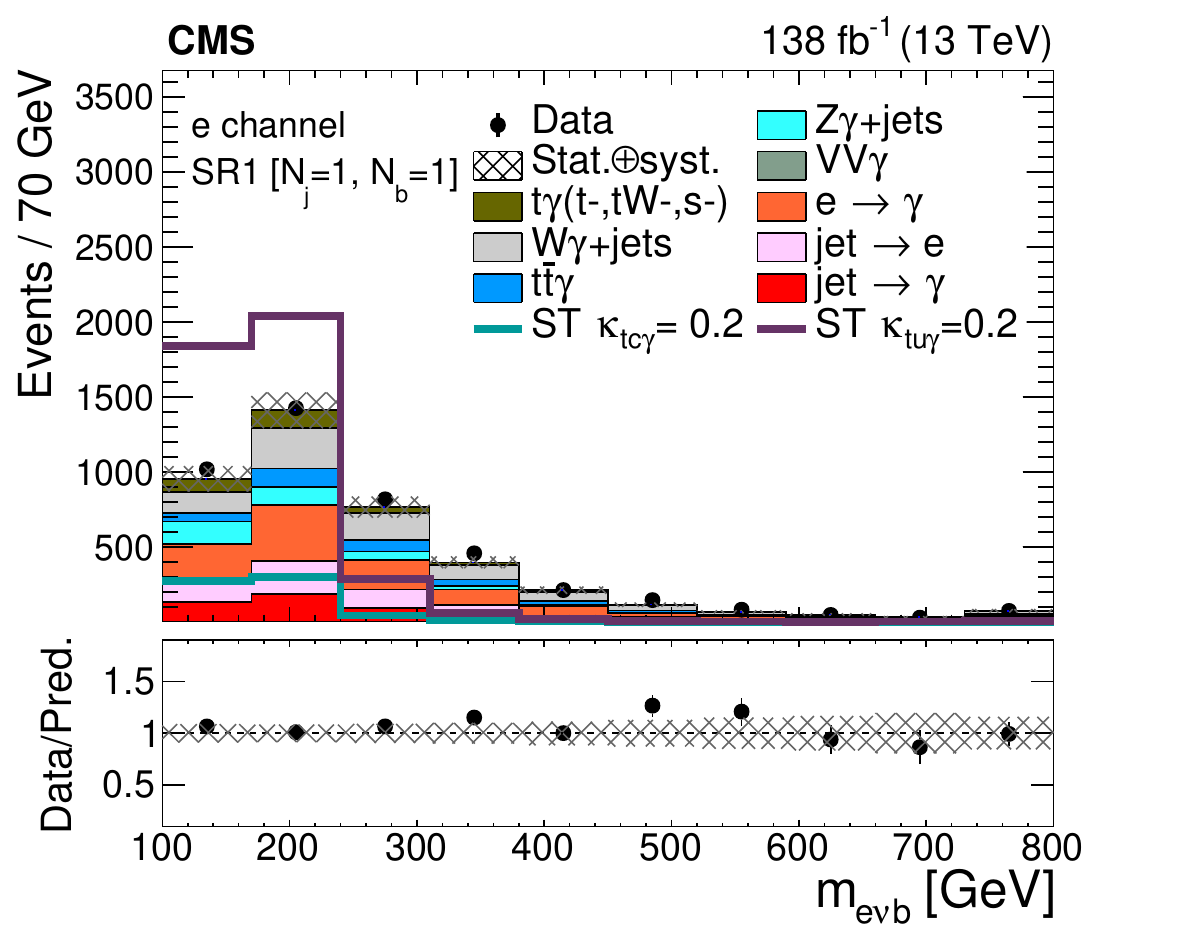}
\includegraphics[width=0.40\textwidth]{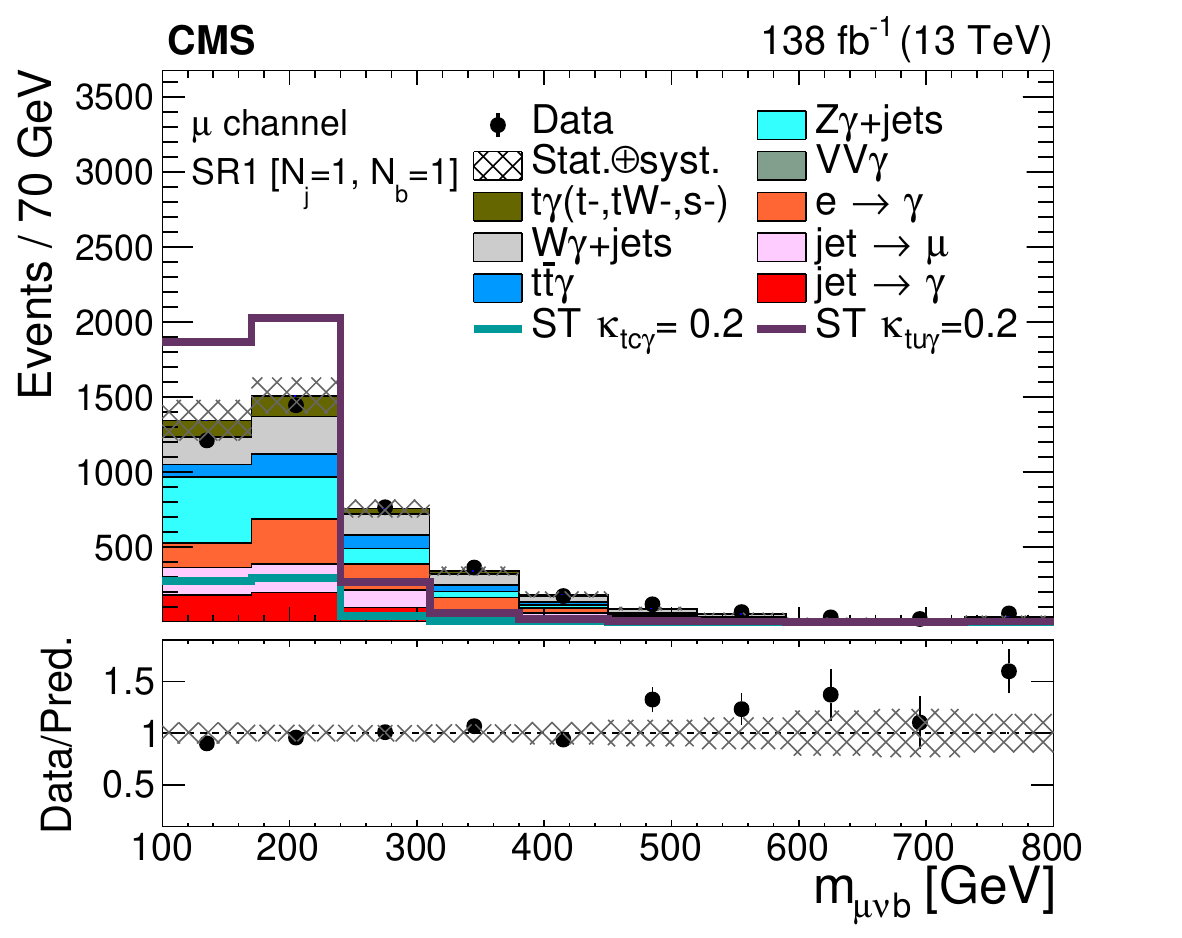}
\includegraphics[width=0.40\textwidth]{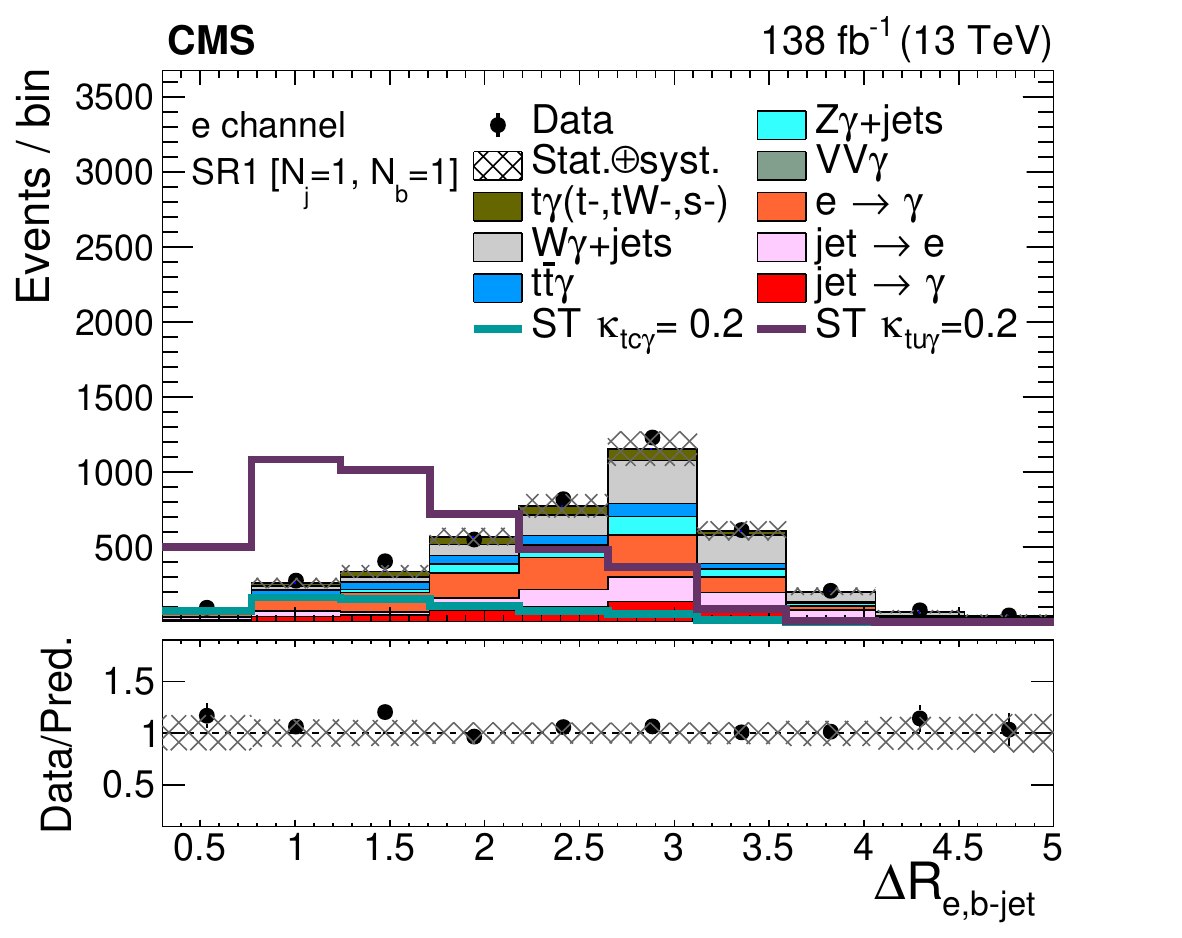}
\includegraphics[width=0.40\textwidth]{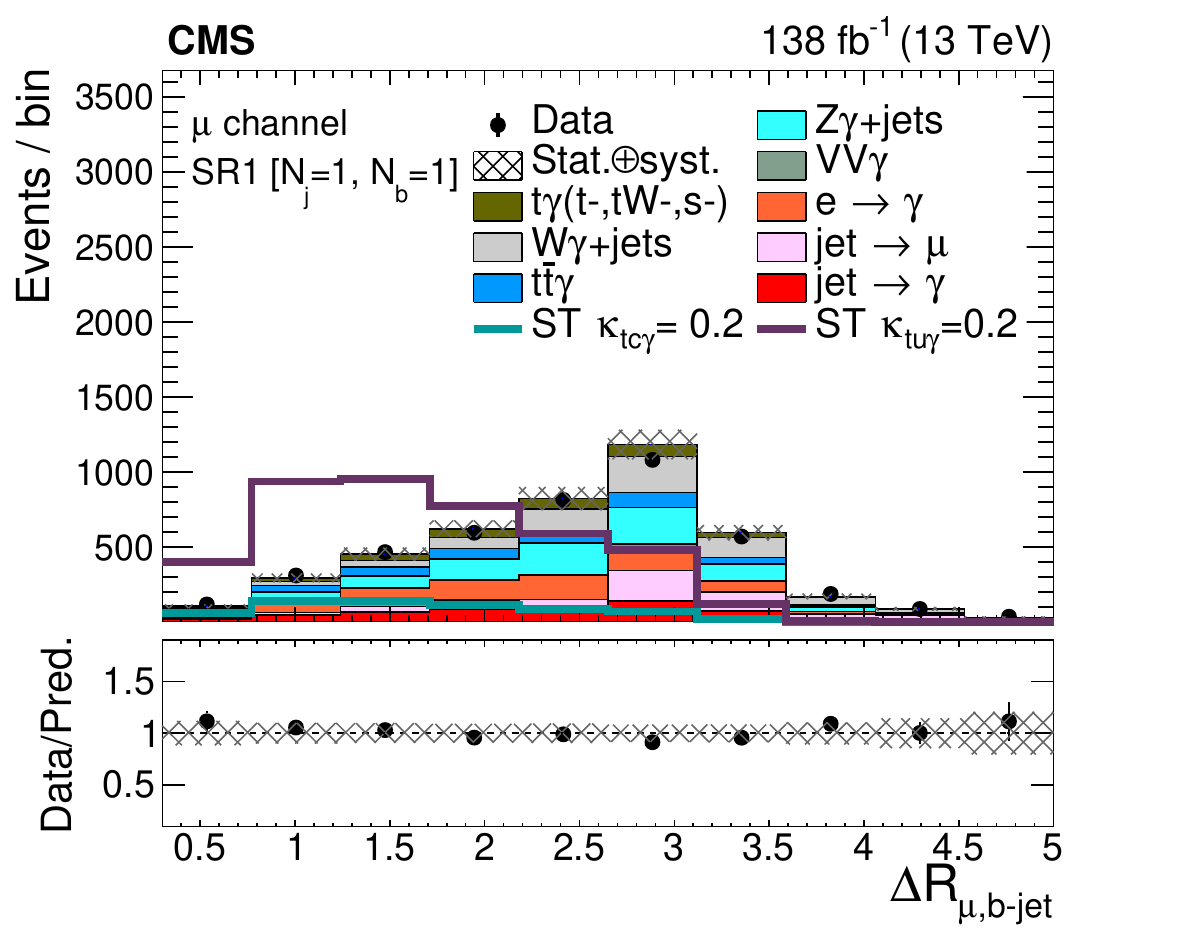}
\caption{From upper to lower, expected and observed distributions of photon
\pt, transverse mass of \PW boson candidate,
reconstructed top quark mass, and \DRlb for the
electron (left) and muon (right) channels in SR1.
For presentational purposes, the \tug and \tcg signal distributions
are normalized to a cross section corresponding to $\kappatug = \kappatcg = 0.2$ and are superimposed on the background expectations.
The last bins include overflows. The vertical bars on the points depict the data statistical uncertainties and the hatched bands show the combined statistical and systematic
uncertainties in the estimated background processes.\label{fig:figsSR1}}
\end{figure*}

\begin{figure*}[!p]
\centering
\includegraphics[width=0.40\textwidth]{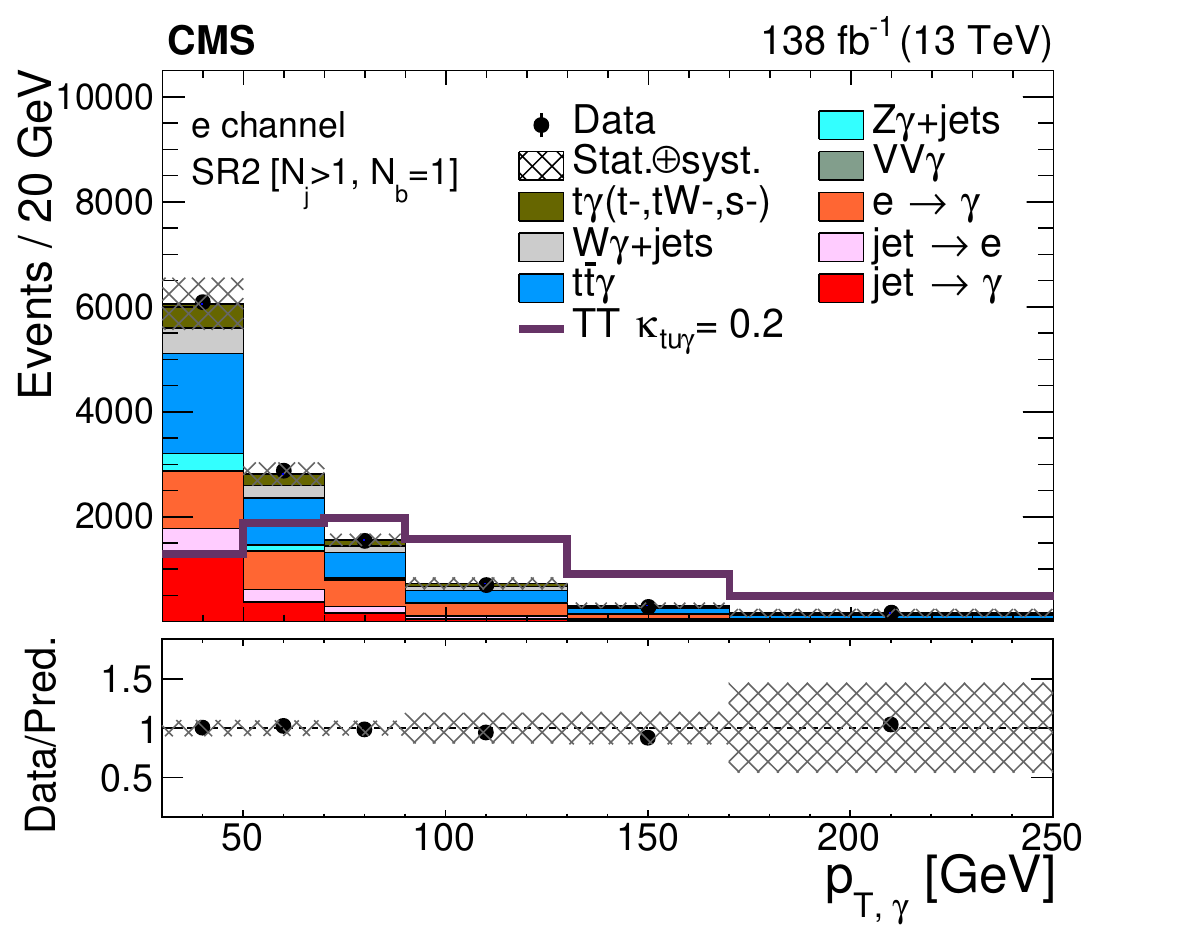}
\includegraphics[width=0.40\textwidth]{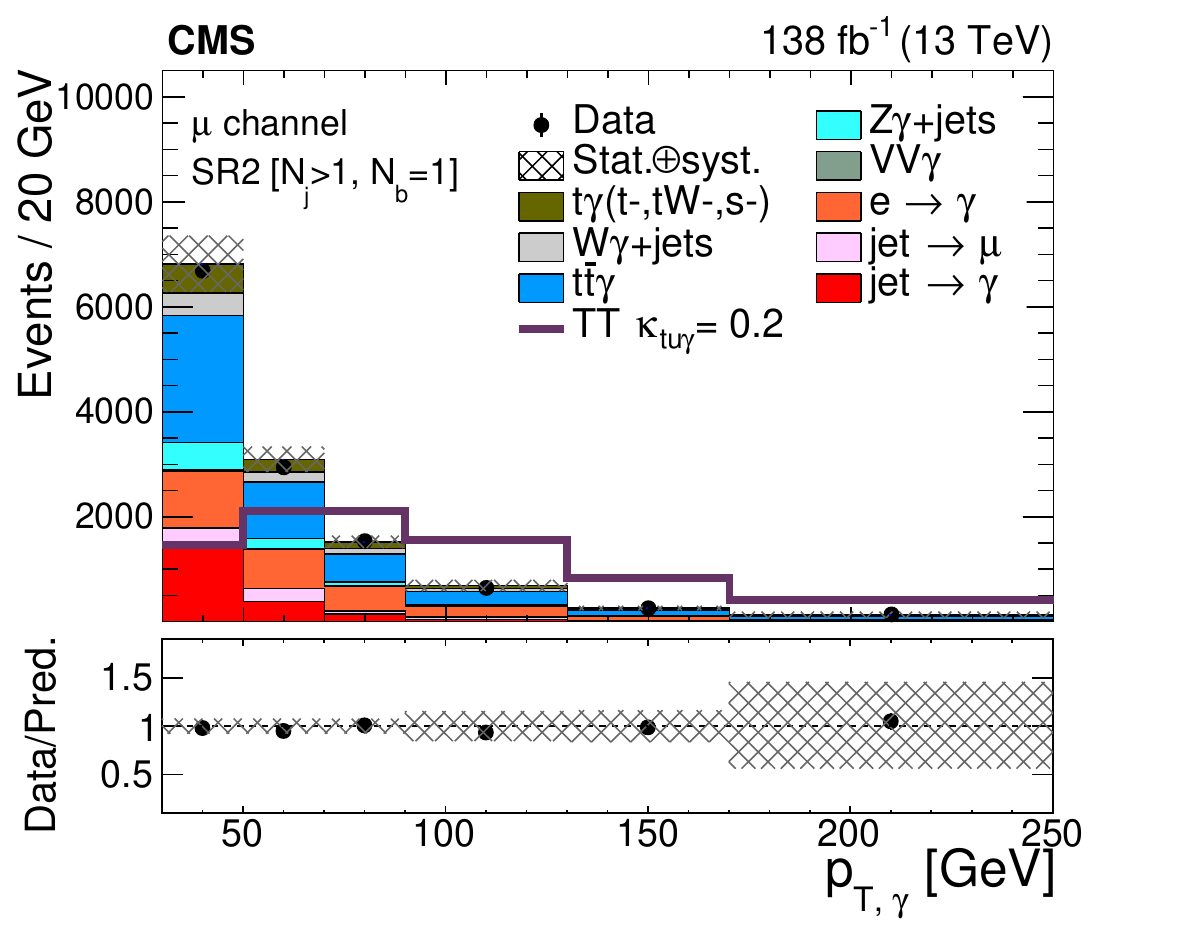}
\includegraphics[width=0.40\textwidth]{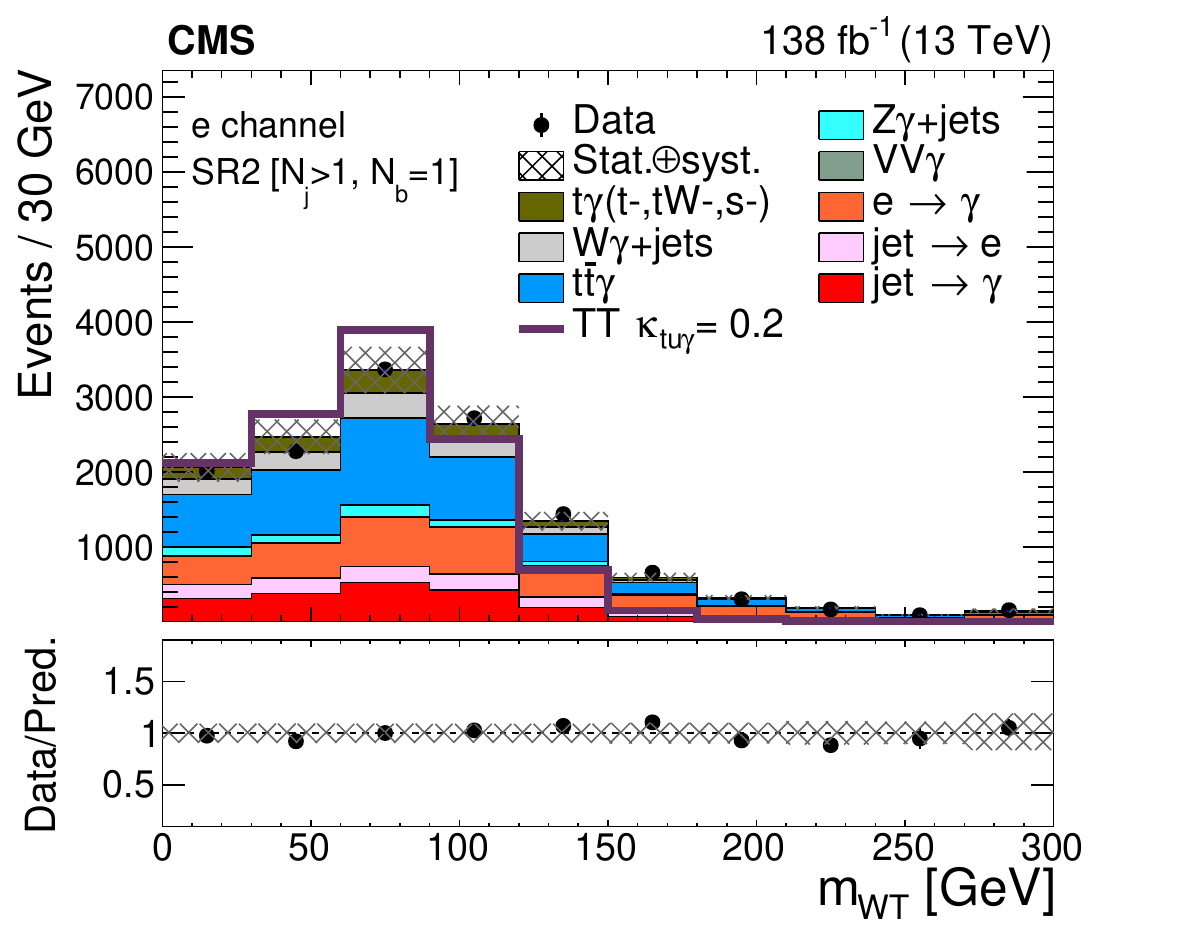}
\includegraphics[width=0.40\textwidth]{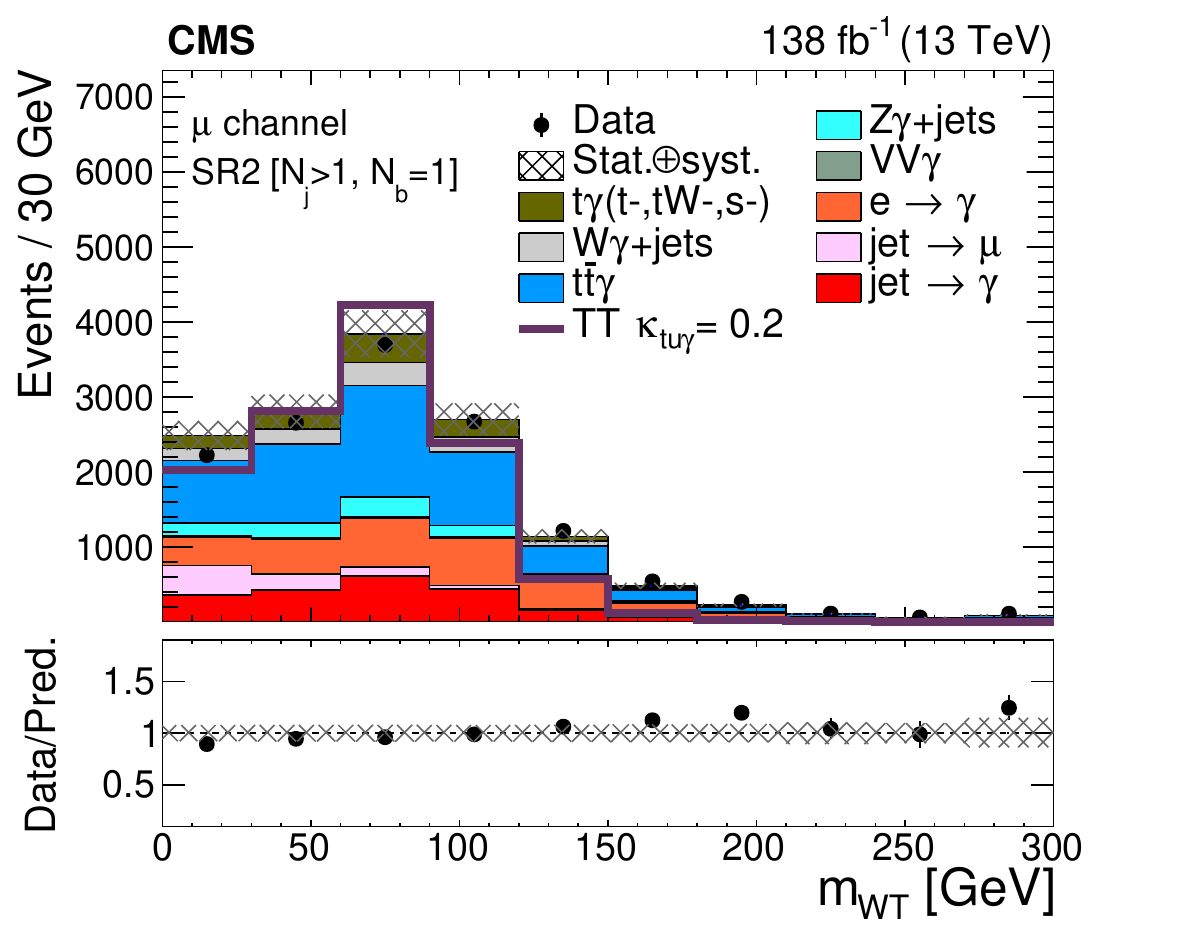}
\includegraphics[width=0.40\textwidth]{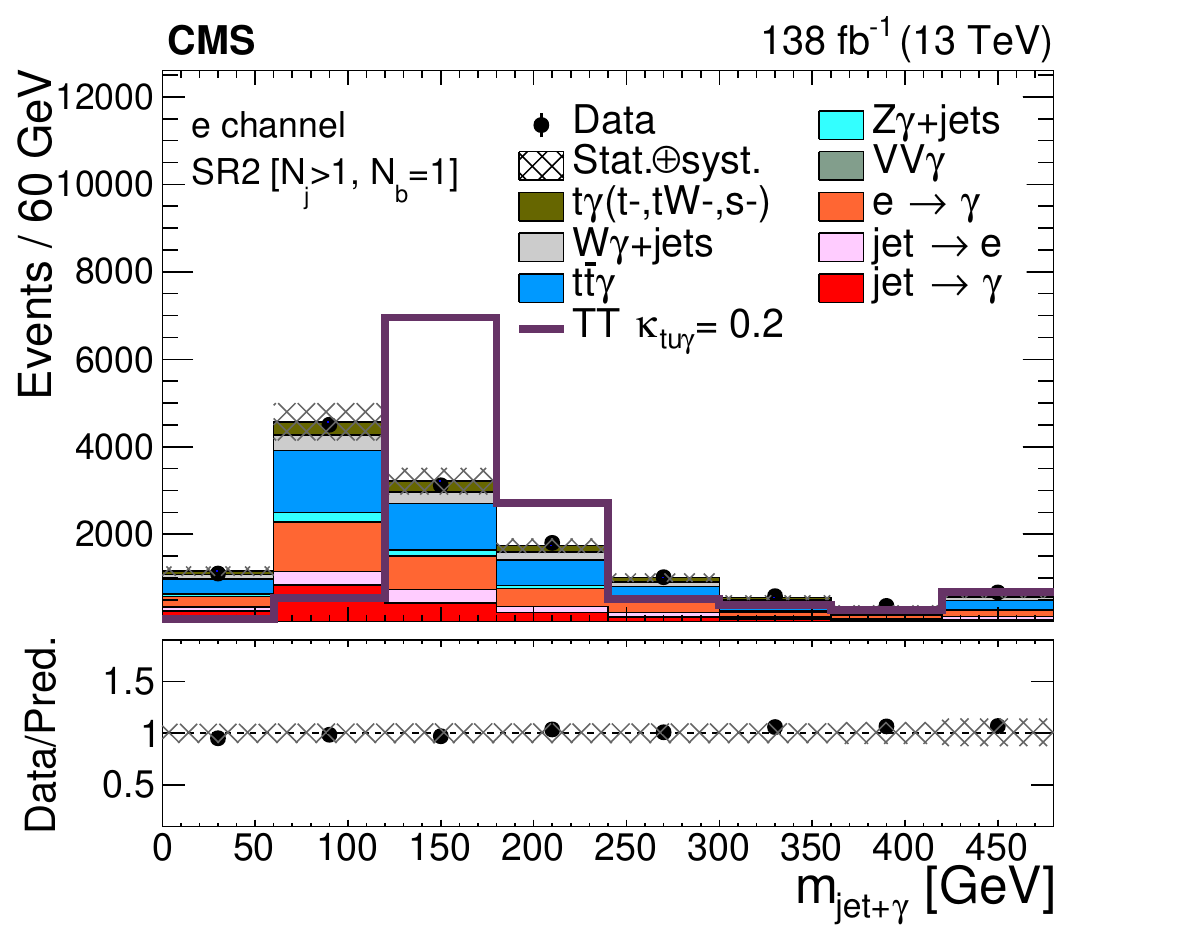}
\includegraphics[width=0.40\textwidth]{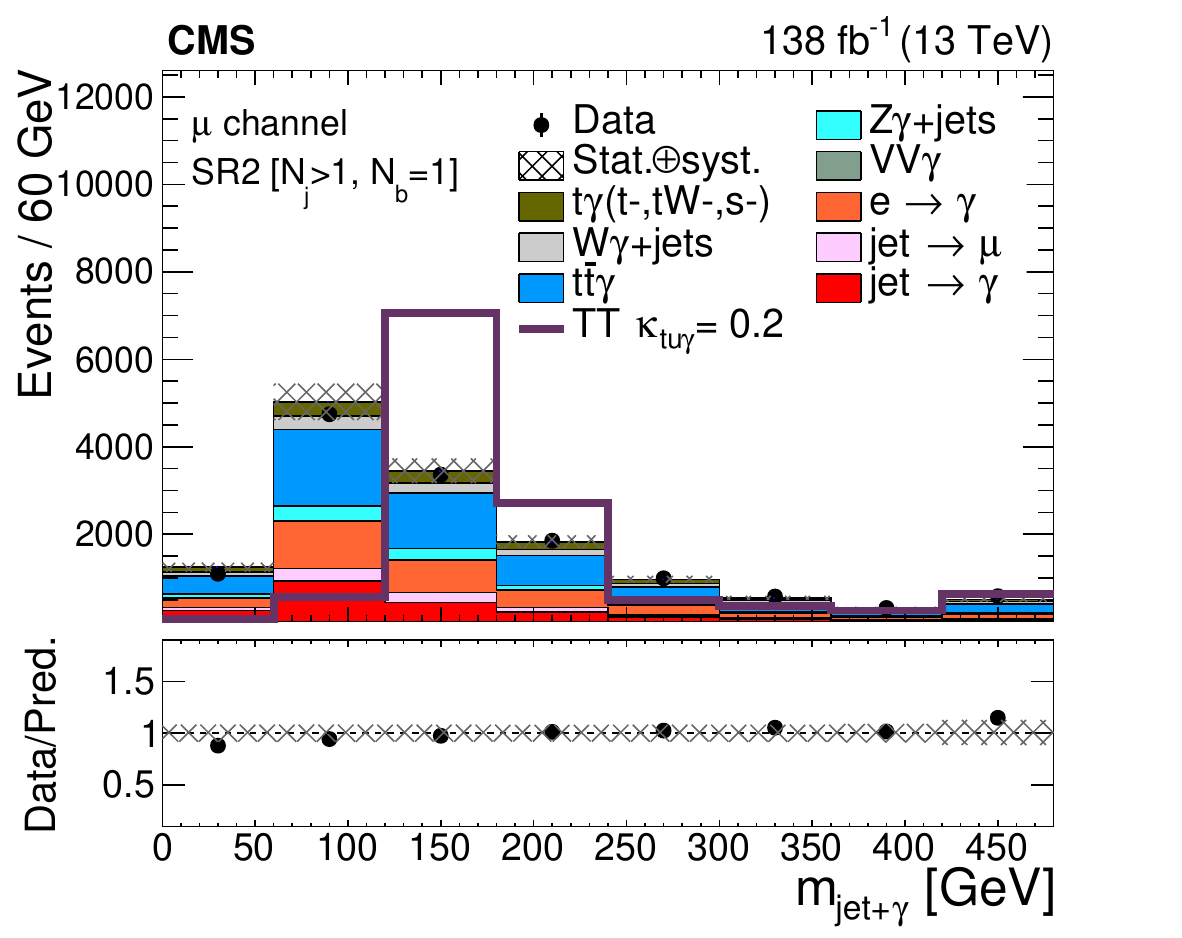}
\includegraphics[width=0.40\textwidth]{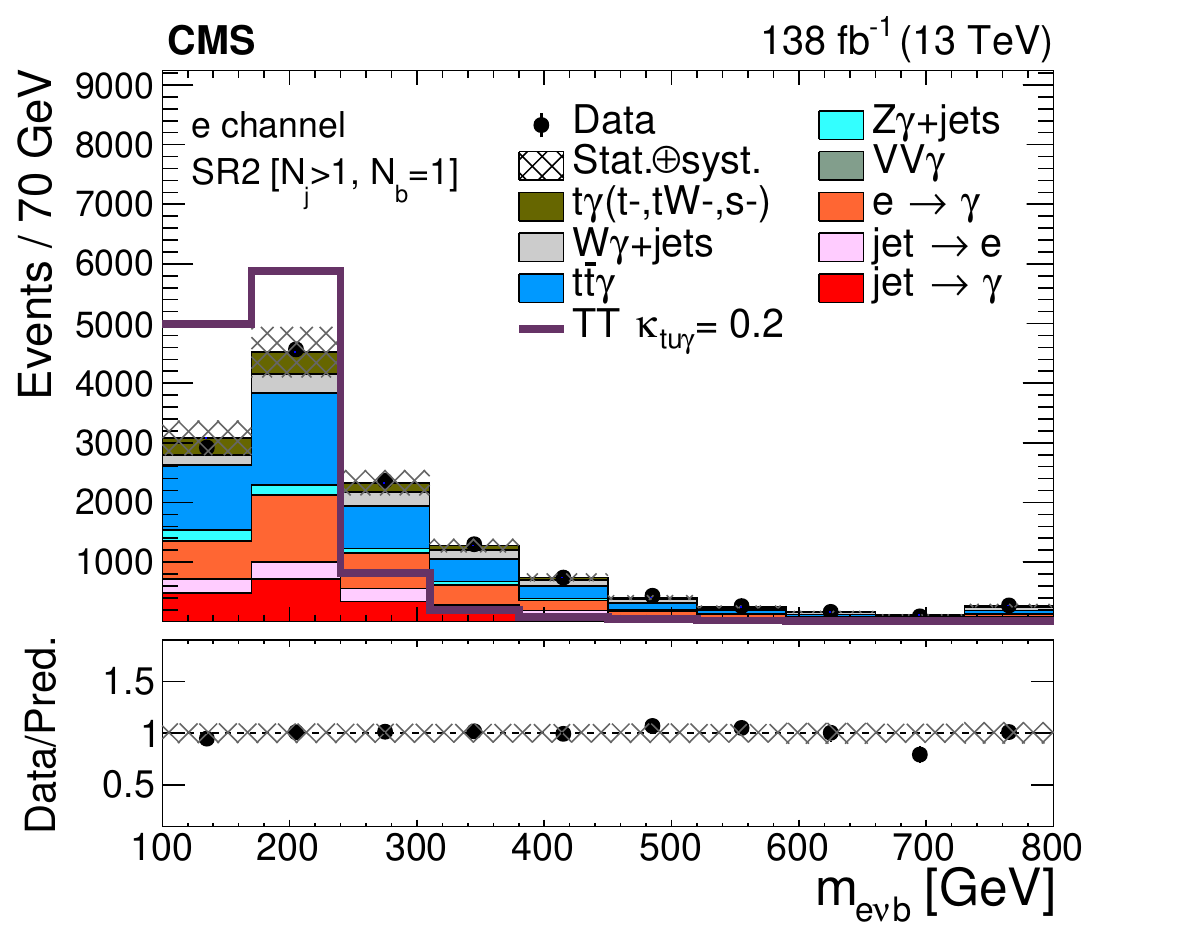}
\includegraphics[width=0.40\textwidth]{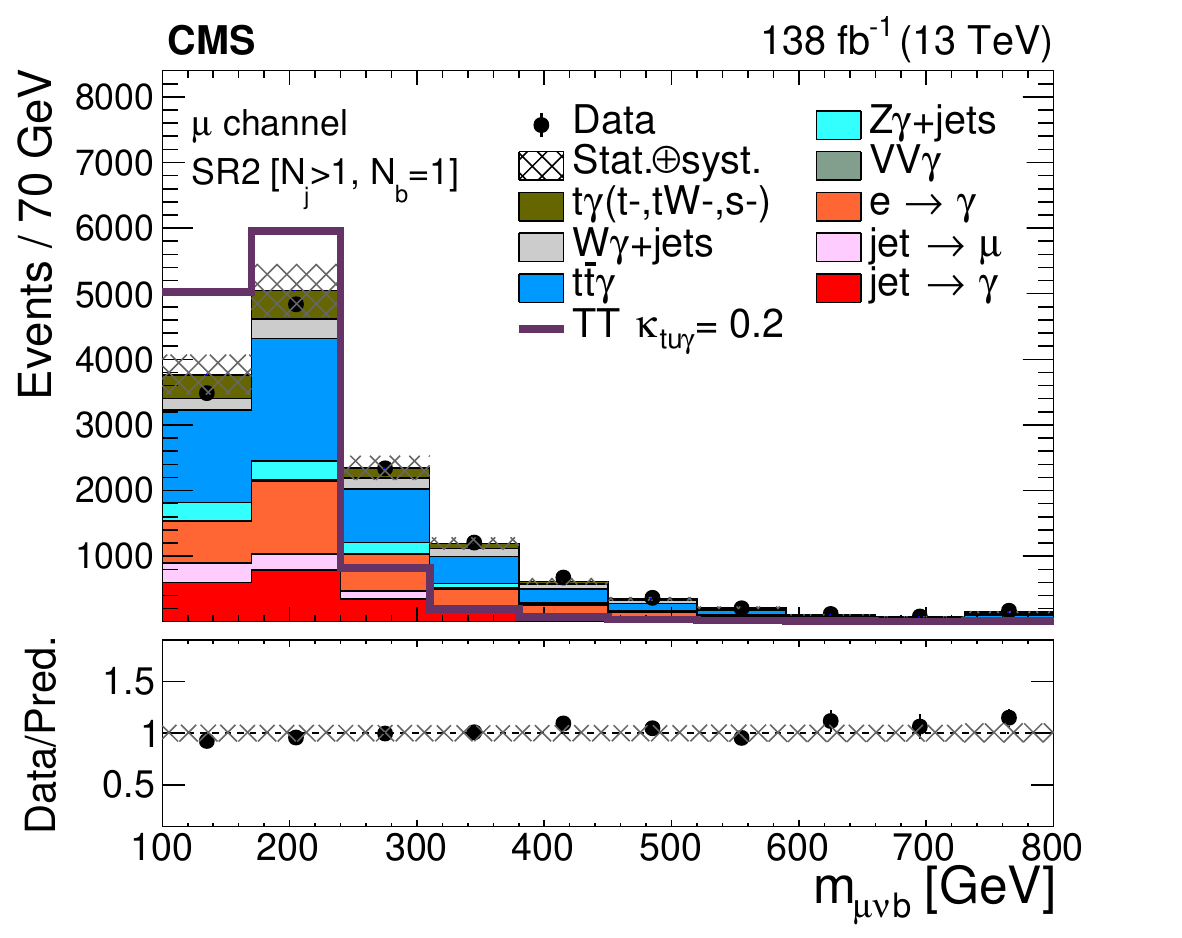}
\caption{From upper to lower, expected and observed distributions of photon \pt, transverse mass of \PW
boson candidate, invariant mass of jet and photon, and reconstructed top quark mass, for the
electron (left) and muon (right) channels in SR2.
For presentational purposes, the \tug signal distributions
are normalized to a cross section corresponding to $\kappatug=0.2$ and are superimposed on the background expectations. The \tcg distributions are not shown as they are the same as the \tug distributions.
The last bins include overflows. The vertical bars on the points depict the data statistical uncertainties and the hatched bands show the combined statistical and systematic
uncertainties in the estimated background processes.\label{fig:figsSR2}}
\end{figure*}

\begin{figure*}[!t]
\centering
\includegraphics[width=0.45\textwidth]{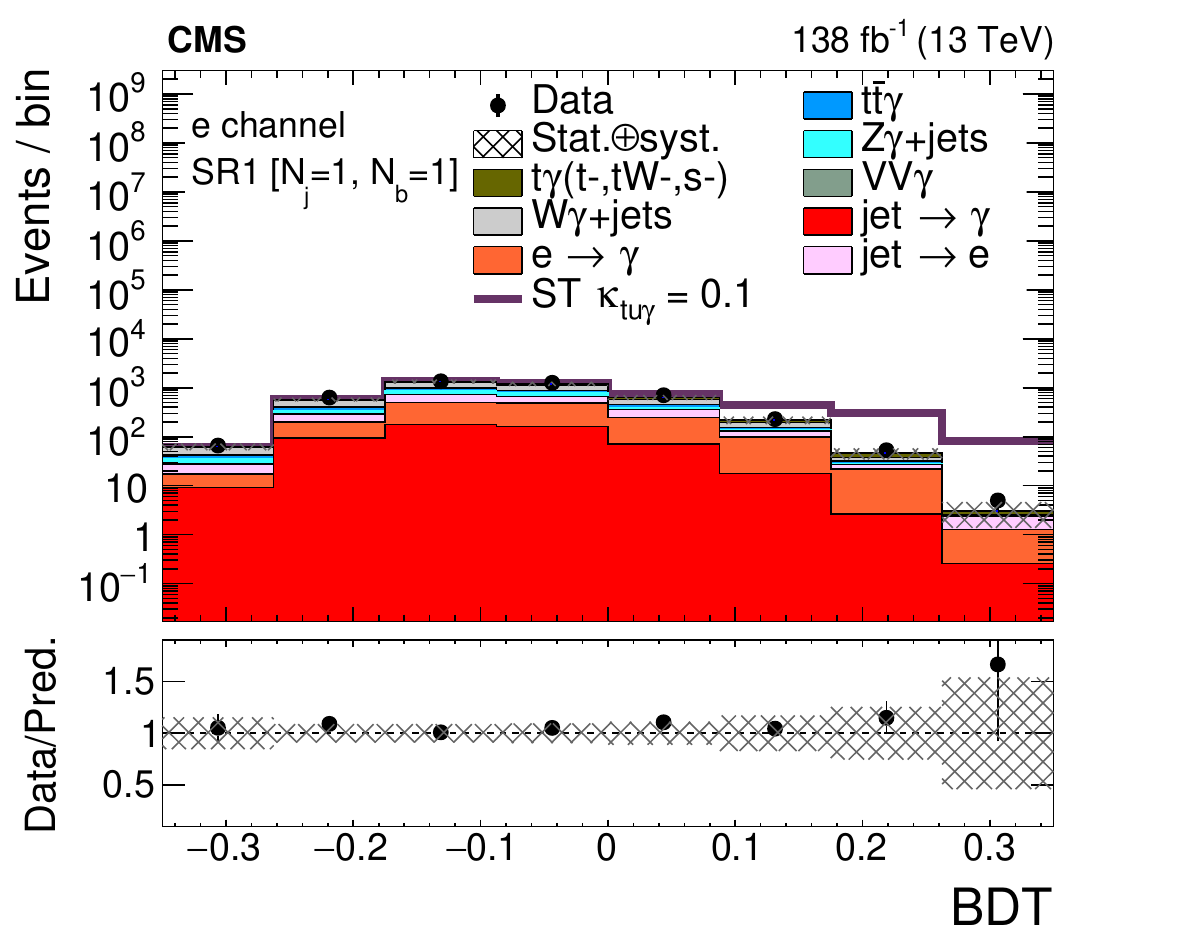}
\includegraphics[width=0.45\textwidth]{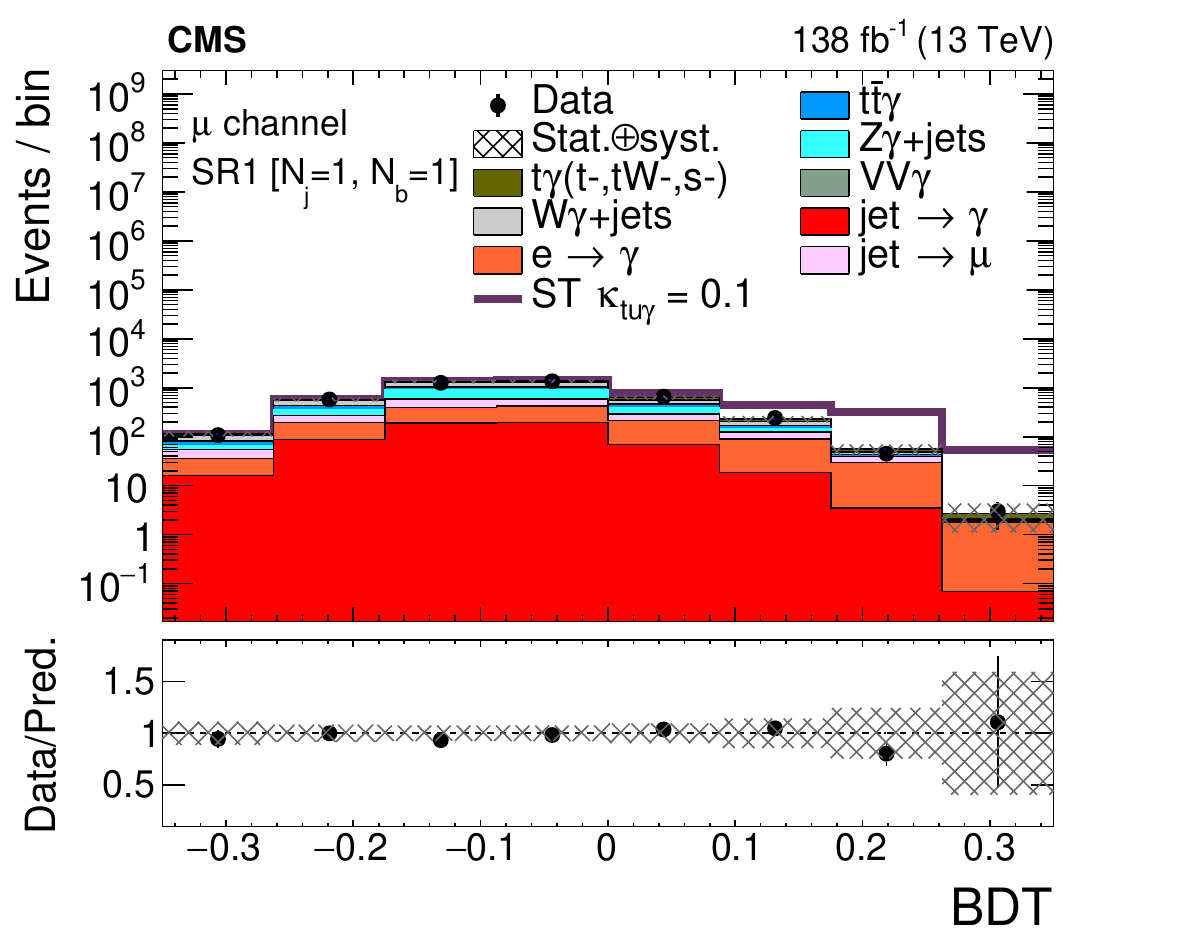}
\includegraphics[width=0.45\textwidth]{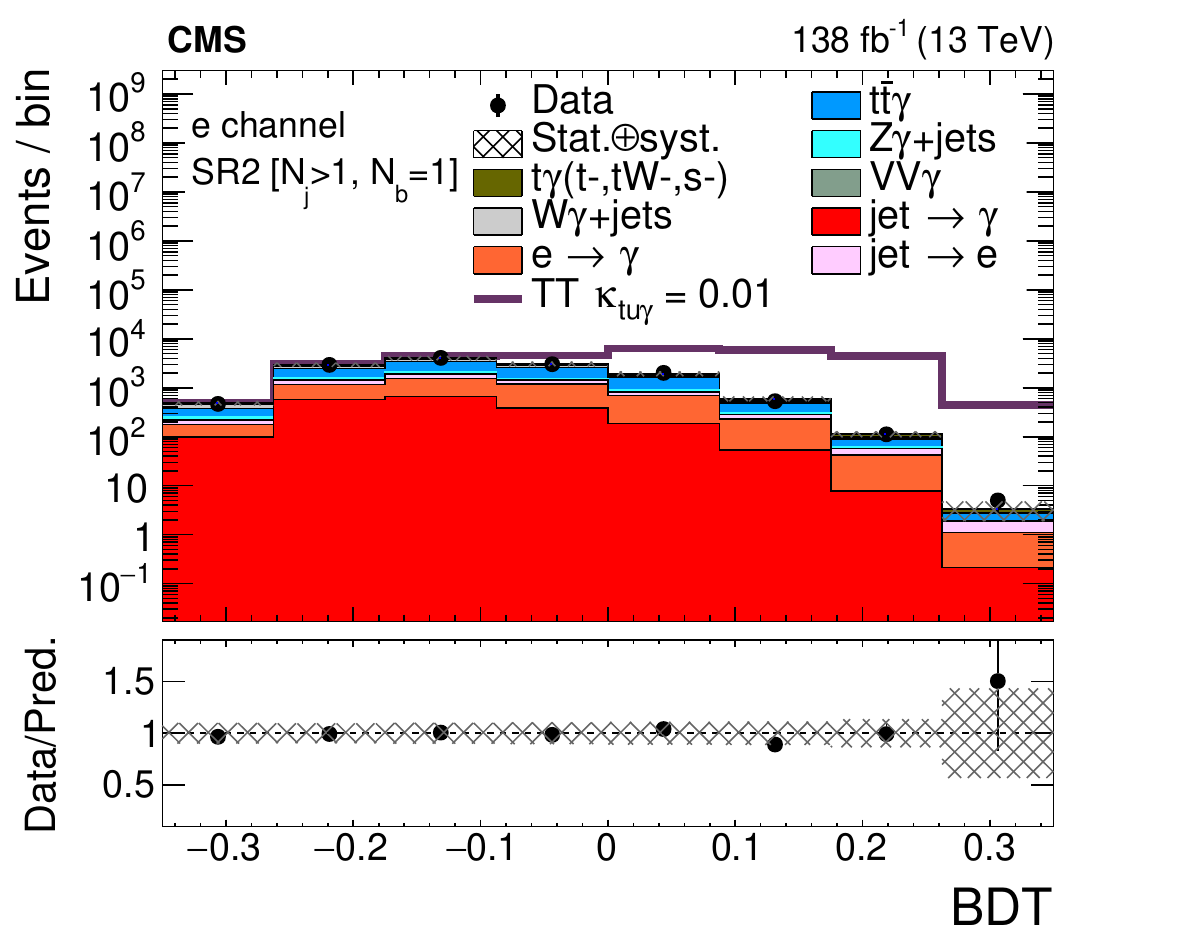}
\includegraphics[width=0.45\textwidth]{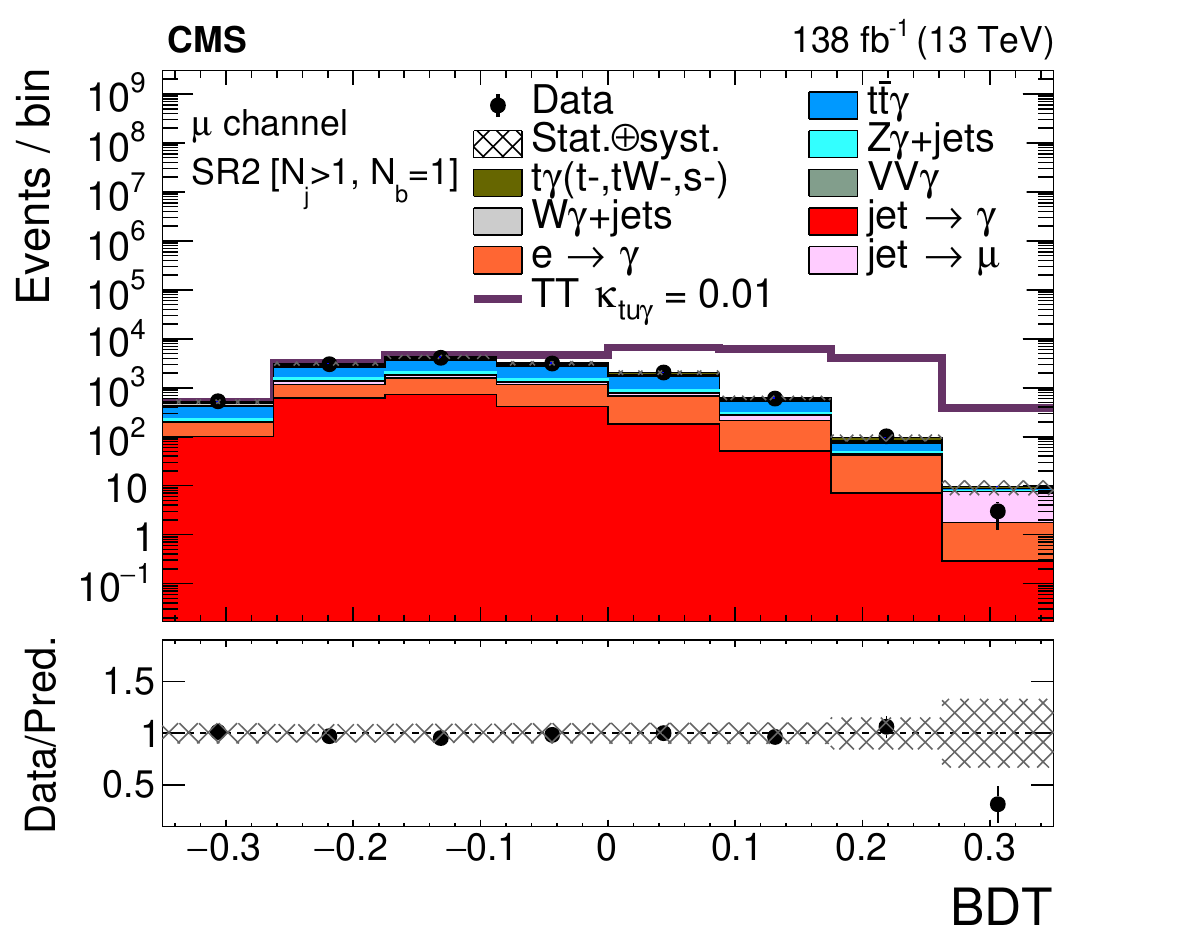}
\caption{The BDT output distributions
for the data, the background predictions, and the expected \tug signal for
electron (left) and muon (right) channels in SR1 (upper) and SR2 (lower).
The signal distribution is normalized to a cross section corresponding to
$\kappatug=0.10$ (0.01) for ST (TT) and is stacked on the background expectations. The first bins include
underflows, and the last bins include overflows. The vertical bars on the points depict the data statistical uncertainties and the hatched bands show the combined statistical and systematic
uncertainties in the estimated background processes.\label{fig:bdtoutputsr1}}
\end{figure*}

\begin{figure*}[!t]
\centering
\includegraphics[width=0.45\textwidth]{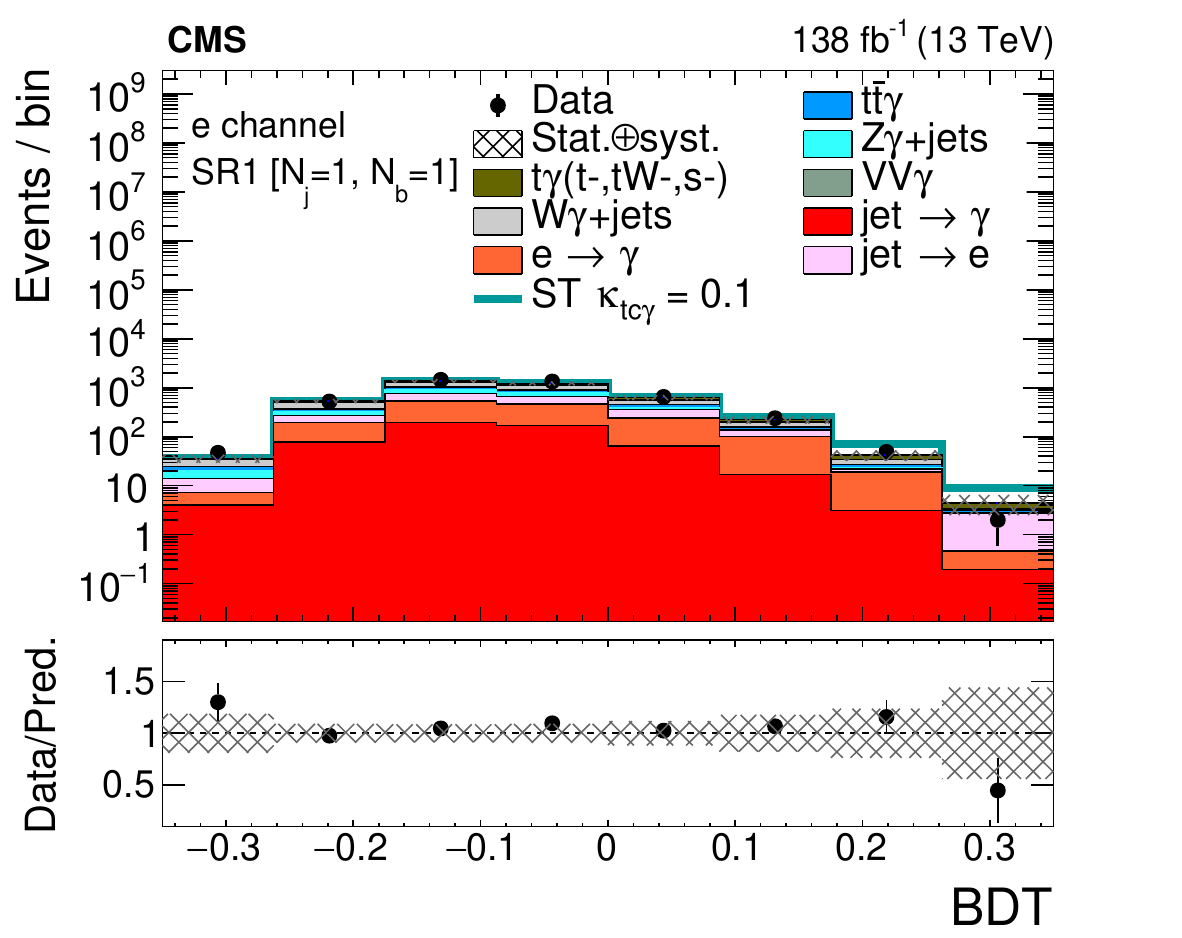}
\includegraphics[width=0.45\textwidth]{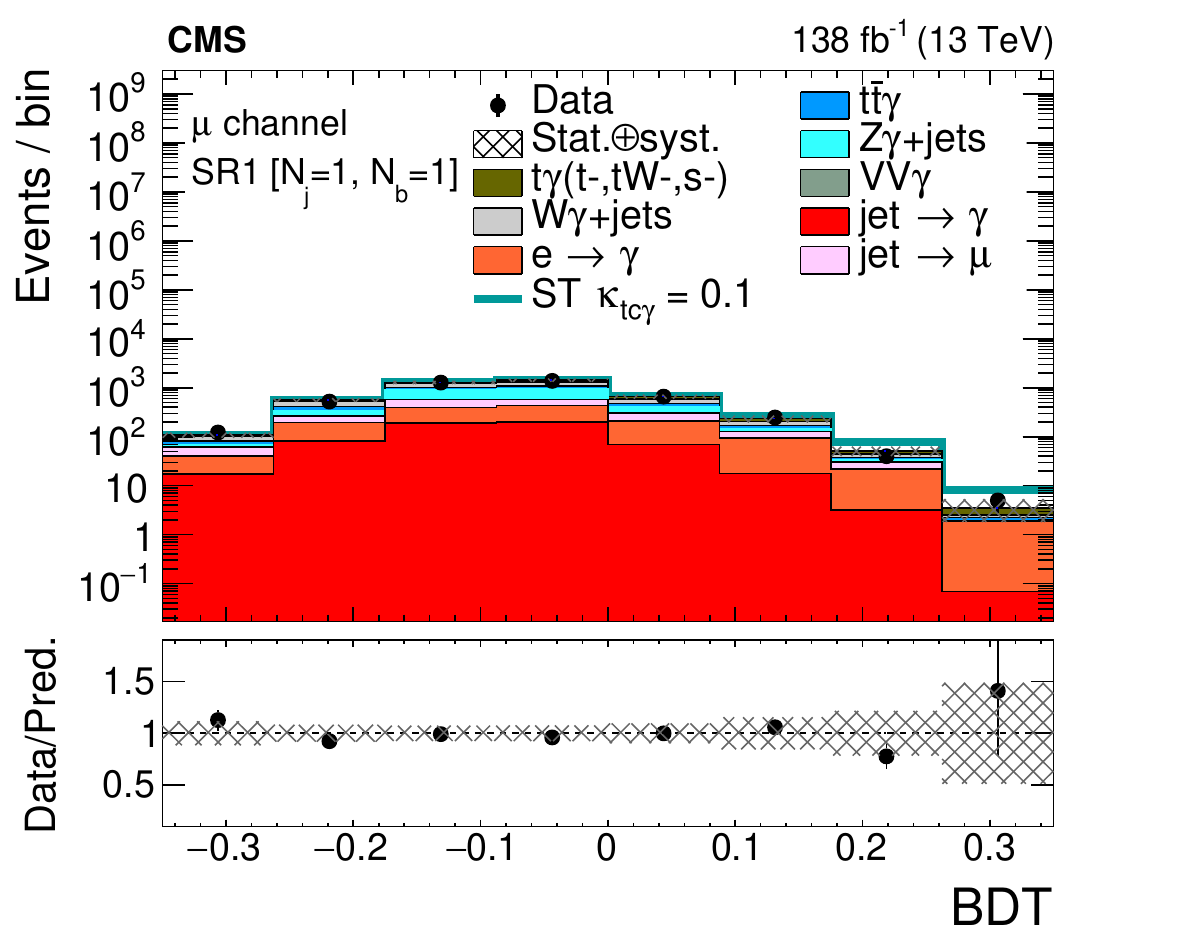}
\includegraphics[width=0.45\textwidth]{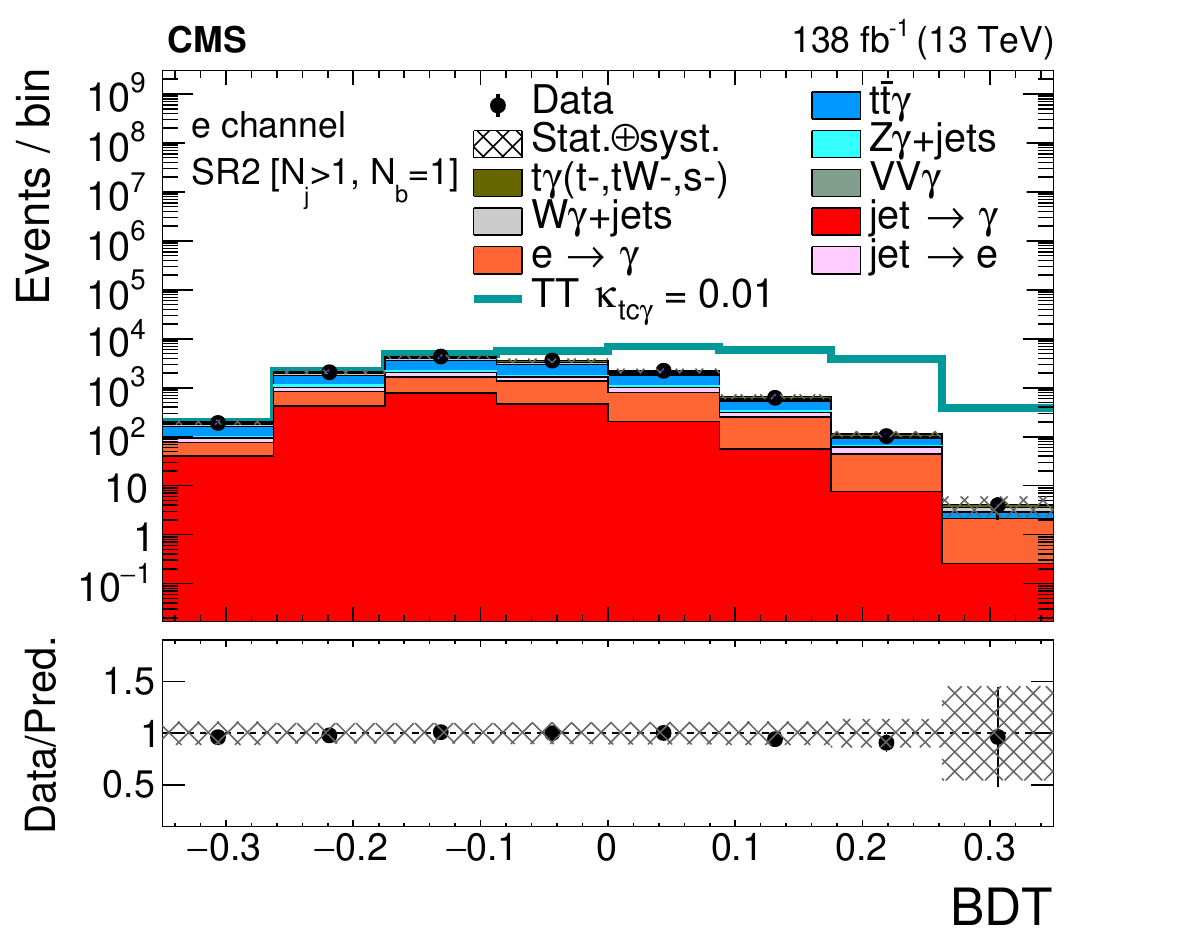}
\includegraphics[width=0.45\textwidth]{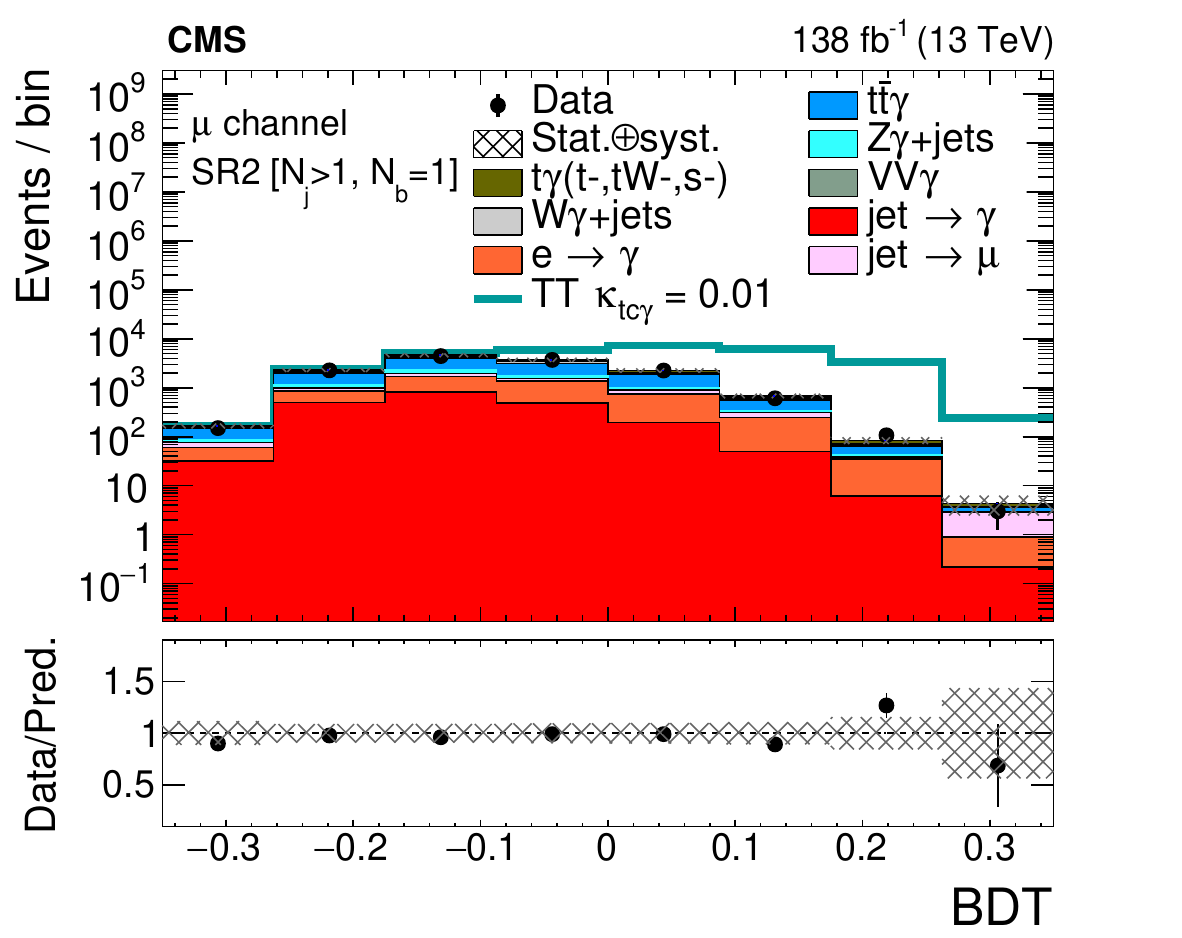}
\caption{The BDT output distributions
for the data, the background predictions, and the expected \tcg signal for
 electron (left) and muon (right) channels in SR1 (upper) and SR2 (lower).
The signal distribution is normalized to a cross section corresponding to
$\kappatcg=0.10$ (0.01) for ST (TT) and is stacked on the background expectations.  The first bins include
underflows, and the last bins include overflows. The vertical bars on the points depict the data statistical uncertainties and the hatched bands show the combined statistical and systematic
uncertainties in the estimated background processes.\label{fig:bdtoutputsr2}}
\end{figure*}

The distributions for four selective discriminating BDT input variables are shown for the predicted
background events and the observed data events from SR1 and SR2 in
Figs.~\ref{fig:figsSR1} and \ref{fig:figsSR2}, respectively. The
distributions from each channel are shown separately and the
background events are estimated as described in Section~\ref{sec:bkg}. The
figures also include the expected contribution of a \tug and
\tcg signal for a coupling of $\kappatqg = 0.2$.
The output BDT distributions of the data, estimated backgrounds, and simulated signals are shown in Figs.~\ref{fig:bdtoutputsr1} and
\ref{fig:bdtoutputsr2} with the BDT trained to select \tug
and \tcg events, respectively, for the combined 2016--2018 data-taking period. Each figure shows the
distributions separated by channel and SRs. For presentational purposes, the simulated
FCNC signals are normalized to $\kappatqg = 0.10$ and 0.01
for SR1 and SR2, respectively. A binned maximum likelihood fit is used to
combine the 12 BDT output distributions from the two channels, two SRs, and three data-taking years,
with systematic uncertainties incorporated as nuisance parameters for \tug and \tcg separately.

\section{Systematic uncertainties}
\label{sec:sys}

We consider an extensive list of systematic uncertainties from
experimental sources, which primarily affect the background estimation,
and from theoretical sources, which primarily affect the signal efficiency.
The estimate of each source of uncertainty is determined by varying the
relevant quantity and evaluating the effect on the
yield and shape of the signal and background BDT distributions.  The
amount of variation is controlled by nuisance parameters in the maximum likelihood
fit.  We also account for correlations among the systematic uncertainties.  In particular,
all uncertainties are correlated between the two SRs.
Uncertainties specific to a lepton flavor are uncorrelated between the two lepton-flavor channels, while
other uncertainties are correlated.  The correlation between the
different data-taking years is described below.

The integrated luminosity is used to normalize predictions obtained from simulation.  The
associated uncertainties for the data collected in 2016, 2017, and 2018
are 1.2, 2.3, and 2.5\%, respectively, and are 30\% correlated~\cite{CMS:2021xjt,CMS:2018elu,CMS:2019jhq}.
The pileup uncertainty affects the distribution of the number of \pp collisions per
bunch crossing. It is estimated by varying the total
\pp inelastic cross section by 4.6\%~\cite{Sirunyan:2020foa} and is fully correlated for all three
years.

Uncertainties in the jet energy scale are derived by recalculating the
four-momentum of each jet using variations from a variety of sources, which are binned in \pt and \sigeta~\cite{Khachatryan:2016kdb}.
The variation is also propagated to \ptvecmiss. The amount of
correlation for each source is evaluated separately across the three years. The uncertainty arising from
the jet energy resolution is estimated by varying the simulated JER by its uncertainty
in bins of \sigeta~\cite{Khachatryan:2016kdb}. This variation is also propagated
to \ptvecmiss. The JER uncertainties are assumed to be
uncorrelated between the three data-taking years. The uncertainty from the calibration of the photon energy
scale is assessed by varying the photon energy by $\pm0.1\%$,
which is obtained from a study of $\PZ\to\EE$ events~\cite{Sirunyan:2020ycc}.

The uncertainty associated with the data-to-simulation scale factors for the
\PQb tagging efficiency for \PQb jets as well as the mistagging efficiencies for \PQc jets and
light-flavor and gluon jets are estimated by varying the scale
factors within their uncertainties measured from a dedicated
\ttbar sample~\cite{CMS:2017wtu}. Each uncertainty is further split into correlated
and uncorrelated components across the three years.

To correct the observed differences in the efficiencies between data
and simulation in lepton and photon reconstruction, identification, and isolation, \pt and \sigeta dependent scale factors with their
relevant uncertainties are derived using the ``tag-and-probe'' method
\cite{Sirunyan:2020ycc,Sirunyan:2018fpa}. The impact of these sources is estimated by
varying the scale factors within their uncertainties, which are
propagated to the final fitting variables.  The statistical component
of these uncertainties is considered uncorrelated between the data-taking years
while the systematic components are assumed to be fully correlated.
The scale factors of leptons are derived from
{\PZ}+jets events but applied to the kinematic region dominated by
\ttbar events, where the lepton isolation efficiency is reduced
due to a larger number of jets. Therefore, to cover the possible
differences of efficiencies between these two kinematic regions, an additional uncertainty of 1\% for the electron channel and
0.5\% for the muon channel is applied~\cite{CMS:2020mpn}.

The trigger efficiency uncertainty is assessed by varying the scale
factors of each trigger within their uncertainties~\cite{Sirunyan:2020ycc}. The resulting
uncertainties are considered uncorrelated among the three years of
data taking. To mitigate the gradual shift of the L1
trigger timing in the forward endcap region ($\abseta < 2.4$) of the ECAL
detector during the years 2016--2017,
a correction factor is applied~\cite{CMS:2020cmk}. The correction factor and its
uncertainty are used to reweight the relevant simulated
events, accounting for the trigger inefficiency. The L1 inefficiency uncertainties are
fully correlated between 2016 and 2017 data samples.

The uncertainty in the normalization of \WWgamma is estimated from
\MGvATNLO to be 23\%.
The normalization uncertainties for the
$\PW\PW$~\cite{CMS:2020mxy}, $\PZ\PZ$~\cite{CMS:2020gtj}, $\PW\PZ$~\cite{Sirunyan:2020gyx}, \tgamma~\cite{ATLAS:2023qdu}, single top quark
\tW-channel~\cite{Sirunyan:2018lcp}, and single top quark $s$-channel~\cite{Kidonakis:2015nna} processes are 5.8, 4.5, 40, 11, 11, and 4\%, respectively,
and are assigned to each background source separately.
The normalization uncertainties are applied in cases where the background normalization is obtained from
simulation.

The systematic uncertainty in the nonprompt-photon background estimation is calculated from three contributions.
The first source is from the definition of the sideband regions in the ABCD
method. To assess this uncertainty, the borders of the sideband regions are changed by 50\% of their nominal values and the EFs are recalculated. The choice of 50\% is made to keep the statistical uncertainty of the new sideband regions under 10\%.
The second source is the potential contribution of prompt-photon events in the sideband regions, which is
not accounted for in the nominal method.  We estimate this effect by subtracting the prompt-photon contribution using simulated events and recalculating the EFs. The third source is the EF statistical uncertainties,
especially for the high-\pt photon bins.  The total nonprompt-photon background uncertainty is calculated as the sum in quadrature of the three
sources and the corresponding systematic uncertainty is evaluated by varying the EFs within the associated
uncertainties.
The uncertainty in the nonprompt-lepton background estimation is 20\%, as derived from the
maximum difference found in the check of the method described in Section~\ref{sec:nonpromptlep}.
The uncertainty in the estimation of the misidentified-photon background
arises from the associated scale factor uncertainties and varies within the 16--22\% range depending on the data-taking year.
The background estimation uncertainties are uncorrelated between the three data-taking years.

Theoretical uncertainties arise from the choice of theoretical
assumptions such as the PDFs and the renormalization and factorization
scales, \muR and \muF, used in the cross section calculations. This affects both the calculation of the cross section and the signal efficiency.
The impact of the uncertainties in the matrix elements of the generators due to
\muR and \muF is obtained by independently varying the scales up and down by
factors of 0.5 and 2.
The uncertainty arising from PDFs used in the simulation is
estimated by reweighting the signal events according to the NNPDF sets~\cite{Butterworth:2015oua}. In addition, the uncertainty
resulting from initial- and final-state radiation in the parton shower
of signal events is estimated by shifting the relevant scales by
a factor of two. All of these uncertainties are correlated across the
three data-taking years.

Among the systematic uncertainties, the integrated luminosity, background
normalization, and nonprompt-lepton background and misidentified-photon background uncertainties
affect only the rate, while the uncertainties from the remaining
systematic sources affect both the shape and normalization of
the BDT output distributions used in the fit.

\section{Results}

The data and SM prediction are in agreement within uncertainties and no excess from
FCNC contributions is observed.
A modified frequentist (\CLs) criterion~\cite{Read:2002hq,Junk:1999kv} in the asymptotic approximation~\cite{Cowan:2010js} is used to
compute the upper limits on the signal cross sections and branching fractions.
A test statistic is defined based on the profile likelihood ratio for the compatibility of the data with background-only and signal+background hypotheses.

\begin{table*}[!t]
\centering
\topcaption{The expected and observed 95\% \CL upper limits using the \CLs criterion on the anomalous couplings \kappatug, \kappatcg
and the corresponding branching fractions \BRtug and \BRtcg from the combination of
the electron and muon channels at NLO for SR1, SR2, and combined (SR1+SR2).
\label{tab:combinations-total}}
\renewcommand{\arraystretch}{1.2}
\begin{scotch}{cccccc}
& & Observed limit & Expected limit & $\pm1\sigma$ (expected limit) & $\pm2\sigma$ (expected limit) \\ \hline
\multirow{4}{*}{\rotatebox{90}{SR1}}
& \kappatug                         &  $12.3\times10^{-3} $  & $11.6\times10^{-3}$    & $(9.7$--$14.4) \times10^{-3}$  &  $(8.1$--$17.4)\times10^{-3}$    \\
& \kappatcg                         &  $15.3\times10^{-3}$   &  $20.1\times10^{-3}$   & $(16.9$--$24.4)\times10^{-3}$   &  $(14.4$--$29.3)\times10^{-3}$    \\
& \BRtug       &  $3.79\times10^{-5}$  & $3.39\times10^{-5}$   & $(2.33$--$5.16)\times10^{-5}$ & $(1.65$--$7.55)\times10^{-5}$    \\
& \BRtcg       &  $5.85\times10^{-5}$  &   $10.11\times10^{-5}$ & $(7.13$--$14.95)\times10^{-5}$ &  $(5.22$--$21.44)\times10^{-5}$   \\[\cmsTabSkip]
\multirow{4}{*}{\rotatebox{90}{SR2}}
& \kappatug                         &  $6.3\times10^{-3} $  & $7.5\times10^{-3}$    & $(6.3$--$9.1) \times10^{-3}$  &  $(5.5$--$11.0)\times10^{-3}$    \\
& \kappatcg                         &  $7.9\times10^{-3}$   &  $8.3\times10^{-3}$   & $(6.8$--$10.0)\times10^{-3}$   &  $(6.0$--$11.8)\times10^{-3}$    \\
& \BRtug       &  $0.98\times10^{-5}$  & $1.41\times10^{-5}$   & $(0.99$--$2.09)\times10^{-5}$ & $(0.75$--$3.02)\times10^{-5}$    \\
& \BRtcg       &  $1.57\times10^{-5}$  &   $1.71\times10^{-5}$ & $(1.14$--$2.52)\times10^{-5}$ &  $(0.89$--$3.51)\times10^{-5}$   \\[\cmsTabSkip]
\multirow{4}{*}{\rotatebox{90}{SR1+SR2}}
& \kappatug                         &  $6.2\times10^{-3} $  & $6.9\times10^{-3}$    & $(5.9$--$8.4) \times10^{-3}$  &  $(5.1$--$10.1)\times10^{-3}$    \\
& \kappatcg                         &  $7.7\times10^{-3}$   &  $7.8\times10^{-3}$   & $(6.7$--$9.7)\times10^{-3}$   &  $(5.7$--$11.5)\times10^{-3}$    \\[\cmsTabSkip]
& \BRtug       &  $0.95\times10^{-5}$  & $1.20\times10^{-5}$   & $(0.89$--$1.78)\times10^{-5}$ & $(0.64$--$2.57)\times10^{-5}$    \\
& \BRtcg       &  $1.51\times10^{-5}$  &   $1.54\times10^{-5}$ & $(1.13$--$2.37)\times10^{-5}$ &  $(0.81$--$3.32)\times10^{-5}$   \\
\end{scotch}
\end{table*}

The limits from the combination of the electron and muon channels
separately for SR1, SR2, and a combination of two SRs are presented
in Table~\ref{tab:combinations-total}.
It summarizes the expected and observed 95\% \CL upper limits on
the anomalous couplings \kappatug and \kappatcg,
and the corresponding branching fractions  \BRtug and \BRtcg with NLO QCD
corrections. The one and two standard deviation ($\pm1\sigma$ and $\pm2\sigma$) ranges of the expected limits are
also presented.
The largest uncertainties affecting the final likelihood fit are the statistical uncertainties from the limited number of simulated events. Among the systematic
uncertainties described in Section~\ref{sec:sys}, the uncertainties due to the nonprompt-photon background, the normalizations of \Zgamma and \Wgamma, and the uncertainty arising from the misidentified-photon background have the greatest impact on the final limits. Overall, the post fit values of nuisance parameters deviate no more than one standard deviation from their initial values.

{\tolerance=800
The results show significant improvements with respect to the previous
CMS results at 8\TeV~\cite{Khachatryan:2015att}. This is due to the
addition of another signal region (SR2) as well as the electron
channel, an increased signal cross section at the higher
center-of-mass energy of 13\TeV, and larger integrated luminosity. 

\section{Summary}

The results of a search for flavor changing neutral current (FCNC) interactions in the top quark sector associated with the \tug and
\tcg vertices have been presented. These vertices are probed
by a simultaneous evaluation of single top quark production in association with a photon and top quark pair production with one
of the top quarks decaying via FCNC. The search is performed using proton-proton
collisions at a center-of-mass energy of 13\TeV, corresponding to an integrated
luminosity of 138\fbinv, collected by the CMS detector at the LHC. The results
are in agreement with the standard model prediction. Upper limits are set at 95\% confidence level on the anomalous
FCNC couplings of $\kappatug < 6.2\times10^{-3}$ and $\kappatcg < 7.7\times10^{-3}$. The
upper limits on the corresponding branching fractions are $\BRtug<0.95\times10^{-5}$ and
$\BRtcg<1.51\times10^{-5}$.
The obtained limit for \BRtug is similar to the current best limit from the ATLAS experiment~\cite{ATLAS:2022per},
while the limit for \BRtcg is
significantly tighter. The result for \BRtcg benefits from the
inclusion of the TT signal, which has comparable sensitivity for
\kappatug and \kappatcg, as well as from having two independent SRs,
one optimized for ST and one for TT.

\begin{acknowledgments}
We congratulate our colleagues in the CERN accelerator departments for the excellent performance of the LHC and thank the technical and administrative staffs at CERN and at other CMS institutes for their contributions to the success of the CMS effort. In addition, we gratefully acknowledge the computing centers and personnel of the Worldwide LHC Computing Grid and other centers for delivering so effectively the computing infrastructure essential to our analyses. Finally, we acknowledge the enduring support for the construction and operation of the LHC, the CMS detector, and the supporting computing infrastructure provided by the following funding agencies: SC (Armenia), BMBWF and FWF (Austria); FNRS and FWO (Belgium); CNPq, CAPES, FAPERJ, FAPERGS, and FAPESP (Brazil); MES and BNSF (Bulgaria); CERN; CAS, MoST, and NSFC (China); MINCIENCIAS (Colombia); MSES and CSF (Croatia); RIF (Cyprus); SENESCYT (Ecuador); MoER, ERC PUT and ERDF (Estonia); Academy of Finland, MEC, and HIP (Finland); CEA and CNRS/IN2P3 (France); SRNSF (Georgia); BMBF, DFG, and HGF (Germany); GSRI (Greece); NKFIH (Hungary); DAE and DST (India); Iran IPM (Iran); SFI (Ireland); INFN (Italy); MSIP and NRF (Republic of Korea); MES (Latvia); LAS (Lithuania); MOE and UM (Malaysia); BUAP, CINVESTAV, CONACYT, LNS, SEP, and UASLP-FAI (Mexico); MOS (Montenegro); MBIE (New Zealand); PAEC (Pakistan); MES and NSC (Poland); FCT (Portugal); MESTD (Serbia); MCIN/AEI and PCTI (Spain); MOSTR (Sri Lanka); Swiss Funding Agencies (Switzerland); MST (Taipei); MHESI and NSTDA (Thailand); TUBITAK and TENMAK (Turkey); NASU (Ukraine); STFC (United Kingdom); DOE and NSF (USA).

\hyphenation{Rachada-pisek} Individuals have received support from the Marie-Curie program and the European Research Council and Horizon 2020 Grant, contract Nos.\ 675440, 724704, 752730, 758316, 765710, 824093, and COST Action CA16108 (European Union); the Leventis Foundation; the Alfred P.\ Sloan Foundation; the Alexander von Humboldt Foundation; the Science Committee, project no. 22rl-037 (Armenia); the Belgian Federal Science Policy Office; the Fonds pour la Formation \`a la Recherche dans l'Industrie et dans l'Agriculture (FRIA-Belgium); the Agentschap voor Innovatie door Wetenschap en Technologie (IWT-Belgium); the F.R.S.-FNRS and FWO (Belgium) under the ``Excellence of Science -- EOS" -- be.h project n.\ 30820817; the Beijing Municipal Science \& Technology Commission, No. Z191100007219010 and Fundamental Research Funds for the Central Universities (China); the Ministry of Education, Youth and Sports (MEYS) of the Czech Republic; the Shota Rustaveli National Science Foundation, grant FR-22-985 (Georgia); the Deutsche Forschungsgemeinschaft (DFG), under Germany's Excellence Strategy -- EXC 2121 ``Quantum Universe" -- 390833306, and under project number 400140256 - GRK2497; the Hellenic Foundation for Research and Innovation (HFRI), Project Number 2288 (Greece); the Hungarian Academy of Sciences, the New National Excellence Program - \'UNKP, the NKFIH research grants K 124845, K 124850, K 128713, K 128786, K 129058, K 131991, K 133046, K 138136, K 143460, K 143477, 2020-2.2.1-ED-2021-00181, and TKP2021-NKTA-64 (Hungary); the Council of Science and Industrial Research, India; the Iran National Science Foundation, Iran; ICSC -- National Research Center for High Performance Computing, Big Data and Quantum Computing, funded by the EU NexGeneration program (Italy); the Latvian Council of Science; the Ministry of Education and Science, project no. 2022/WK/14, and the National Science Center, contracts Opus 2021/41/B/ST2/01369 and 2021/43/B/ST2/01552 (Poland); the Funda\c{c}\~ao para a Ci\^encia e a Tecnologia, grant CEECIND/01334/2018 (Portugal); the National Priorities Research Program by Qatar National Research Fund; MCIN/AEI/10.13039/501100011033, ERDF ``a way of making Europe", and the Programa Estatal de Fomento de la Investigaci{\'o}n Cient{\'i}fica y T{\'e}cnica de Excelencia Mar\'{\i}a de Maeztu, grant MDM-2017-0765 and Programa Severo Ochoa del Principado de Asturias (Spain); the Chulalongkorn Academic into Its 2nd Century Project Advancement Project, and the National Science, Research and Innovation Fund via the Program Management Unit for Human Resources \& Institutional Development, Research and Innovation, grant B37G660013 (Thailand); the Kavli Foundation; the Nvidia Corporation; the SuperMicro Corporation; the Welch Foundation, contract C-1845; and the Weston Havens Foundation (USA).
\end{acknowledgments}

\bibliography{auto_generated}
\cleardoublepage \appendix\section{The CMS Collaboration \label{app:collab}}\begin{sloppypar}\hyphenpenalty=5000\widowpenalty=500\clubpenalty=5000\input{TOP-21-013-public-authorlist.tex}\end{sloppypar}
\end{document}

%% file: TOP-21-013-public-authorlist.tex
\cmsinstitute{Yerevan Physics Institute, Yerevan, Armenia}
{\tolerance=6000
A.~Hayrapetyan, A.~Tumasyan\cmsAuthorMark{1}\cmsorcid{0009-0000-0684-6742}
\par}
\cmsinstitute{Institut f\"{u}r Hochenergiephysik, Vienna, Austria}
{\tolerance=6000
W.~Adam\cmsorcid{0000-0001-9099-4341}, J.W.~Andrejkovic, T.~Bergauer\cmsorcid{0000-0002-5786-0293}, S.~Chatterjee\cmsorcid{0000-0003-2660-0349}, K.~Damanakis\cmsorcid{0000-0001-5389-2872}, M.~Dragicevic\cmsorcid{0000-0003-1967-6783}, A.~Escalante~Del~Valle\cmsorcid{0000-0002-9702-6359}, P.S.~Hussain\cmsorcid{0000-0002-4825-5278}, M.~Jeitler\cmsAuthorMark{2}\cmsorcid{0000-0002-5141-9560}, N.~Krammer\cmsorcid{0000-0002-0548-0985}, D.~Liko\cmsorcid{0000-0002-3380-473X}, I.~Mikulec\cmsorcid{0000-0003-0385-2746}, J.~Schieck\cmsAuthorMark{2}\cmsorcid{0000-0002-1058-8093}, R.~Sch\"{o}fbeck\cmsorcid{0000-0002-2332-8784}, D.~Schwarz\cmsorcid{0000-0002-3821-7331}, M.~Sonawane\cmsorcid{0000-0003-0510-7010}, S.~Templ\cmsorcid{0000-0003-3137-5692}, W.~Waltenberger\cmsorcid{0000-0002-6215-7228}, C.-E.~Wulz\cmsAuthorMark{2}\cmsorcid{0000-0001-9226-5812}
% ...
\par}
\vskip\cmsinstskip
\dag:~Deceased\\
$^{1}$Also at Yerevan State University, Yerevan, Armenia\\
$^{2}$Also at TU Wien, Vienna, Austria\\
% ...
